\documentclass[12pt]{article}

\usepackage[dvipsname]{xcolor}

\usepackage{graphicx}
\usepackage{amsmath}        
\usepackage{amssymb}        
\usepackage{slashed}
\usepackage{bm}

\def\mydate{January 18,  2020}
\def\ignore#1{{}}

\def\go{\rightarrow}
\def\dd{\partial}
\def\tr{{\rm tr}\,}
\def\ep{{\epsilon}}
\def\eps{{\varepsilon}}

\def\SM{{\rm SM}}
\def\KK{{\rm KK}}
\def\EM{{\rm EM}}
\def\onehalf{\hbox{$\frac{1}{2}$}}
\def\la{\langle}
\def\ra{\rangle}

\def\mybig{\displaystyle \strut }

\def\myfrac#1#2{\frac{\mybig #1}{\mybig #2}}

\def\mymat#1#2{\begin{matrix}#1 \cr \noalign{\kern -2pt} #2\end{matrix}}

\def\mynoalign{\noalign{\kern 4pt}}
\def\mysnoalign{\noalign{\kern 3pt}}

\makeatletter
\@addtoreset{equation}{section}
\makeatother

\begin{document}

\thispagestyle{empty}

{\small \noindent \mydate    \hfill OU-HET 1022}

{\small \noindent    \hfill KUNS-2771}

\vskip 3.cm

\baselineskip=30pt plus 1pt minus 1pt

\begin{center}
{\Large \bf  CKM matrix and FCNC suppression}

{\Large \bf   in  $\bm{SO(5)\times U(1) \times SU(3)}$ gauge-Higgs unification}

\end{center}


\baselineskip=22pt plus 1pt minus 1pt

\vskip 1.5cm

\begin{center}
{\bf Shuichiro Funatsu$^1$, Hisaki Hatanaka$^2$,  Yutaka Hosotani$^3$,}

{\bf Yuta Orikasa$^4$ and Naoki Yamatsu$^5$}

\baselineskip=18pt plus 1pt minus 1pt

\vskip 10pt
{\small \it $^1$Institute of Particle Physics and Key Laboratory of Quark and Lepton 
Physics (MOE), Central China Normal University, Wuhan, Hubei 430079, China} \\
{\small \it $^2$Osaka, Osaka 536-0014, Japan} \\
{\small \it $^3$Department of Physics, Osaka University, 
Toyonaka, Osaka 560-0043, Japan} \\
{\small \it $^4$Institute of Experimental and Applied Physics, Czech Technical University in Prague,} \\
{\small \it Husova 240/5, 110 00 Prague 1, Czech Republic} \\
{\small \it $^5$Department of Physics, Kyoto University, Kyoto 606-8502, Japan} \\

\end{center}

\vskip 1.5cm
\baselineskip=18pt plus 1pt minus 1pt

\begin{abstract}
The Cabibbo-Kobayashi-Maskawa (CKM) mixing matrix and 
flavor-changing neutral currents (FCNC's)  in the quark sector  are examined in the GUT inspired 
$SO(5)\times U(1) \times SU(3)$ gauge-Higgs unification in which 
the 4D Higgs boson is identified with the Aharonov-Bohm phase in the fifth dimension.
Gauge invariant brane interactions play an important role for
 the flavor mixing in the charged-current weak interactions.
The CKM  matrix is reproduced except that  the up quark mass needs to be  
larger than the observed one.
FCNC's are naturally suppressed as a consequence of
the gauge invariance, with a  factor of order $10^{-6}$.
It is also shown that induced flavor-changing Yukawa couplings are extremely small.
\end{abstract}


\newpage

\baselineskip=20pt plus 1pt minus 1pt
\parskip=0pt

\section{Introduction} 

The standard model (SM), $SU(3)_C \times SU(2)_L \times U(1)_Y$ gauge theory, has been 
firmly established at low energies.   Yet it is not clear what the observed Higgs boson is. 
All of the  Higgs couplings to other fields and to itself  need to be 
determined with better accuracy in the coming experiments.
The fundamental problem is the lack of a principle which regulates the Higgs interactions.

One possible answer is the gauge-Higgs unification  in which the Higgs boson is 
identified with the zero mode of the fifth dimensional component of the
gauge potential.\cite{Hosotani1983}-\cite{Kubo2002}
It appears as a fluctuation mode of the Aharonov-Bohm  (AB) phase 
$\theta_H$  in  the fifth dimension.
As a concrete model,  the $SU(3)_C \times SO(5) \times U(1)_X$ gauge theory in the 
Randall-Sundrum (RS)  warped space has been proposed.\cite{ACP2005}-\cite{FHHOS2013}
It gives nearly the same phenomenology at low energies  as the standard 
model (SM).\cite{FHHOS2013}-\cite{FHHO2017}
Deviations of the gauge couplings of quarks and leptons from the SM values are
less than $10^{-3}$ for $\theta_H \sim 0.1$. 
Higgs couplings of quarks, leptons, $W$ and $Z$ are approximately 
the SM values times $\cos \theta_H$, the deviation being about 1{\%}.
The Kaluza-Klein (KK) mass scale is about $m_\KK \sim 8\,$TeV  for $\theta_H \sim 0.1$. 
The KK excited states contribute, for instance, in intermediate states of the two $\gamma$ decay 
of the Higgs boson. Their contribution is  finite and very small.  
The signal strengths of various Higgs decay modes are approximately $\cos^2 \theta_H$ times 
the SM values.  The branching fractions of those decay modes are  approximately the same as 
in the SM.

Gauge-Higgs unification predicts $Z'$ bosons, which are the first KK modes of $\gamma$, $Z$, and
$Z_R$ ($SU(2)_R$ gauge boson).  Their masses are in the $6\,$TeV-$9\,$TeV range for 
$\theta_H=0.11$-0.07 in the model with quark-lepton multiplets introduced in the 
vector representation of $SO(5)$, which will be referred to as the A-model below.
Those $Z'$ bosons have broad widths and can be produced at 14$\,$TeV LHC.
The current non-observation of $Z'$ signals puts the limit $\theta_H \lesssim 0.11$.
Recently an alternative model with quark-lepton multiplets  introduced in the 
spinor, vector, and singlet representations of $SO(5)$ (referred to as the B-model below) 
has been proposed,\cite{FHHOY2019a} which can be 
incorporated in the $SO(11)$ gauge-Higgs grand unification.\cite{HosotaniYamatsu2015, Furui2016}
Other variants of the fermion content have also been proposed.\cite{Yoon2019a}
Implications of gauge-Higgs unification to precision electroweak observables have been
investigated.   It has been shown that the typical models  are consistent
with the current measurements.

Distinct signals of the gauge-Higgs unification can be found in $e^+ e^-$ 
collisions.\cite{FHHO2017ILC}-\cite{YH2019a}
Large parity violation appears in the couplings of quarks and leptons to
KK gauge bosons, particularly to the $Z'$ bosons.  In the A-model  
right-handed quarks and charged leptons have rather large couplings to $Z'$.
The interference effects of $Z'$ bosons can be
clearly observed at 250$\,$GeV $e^+ e^-$ international linear collider (ILC).
In the process $e^+ e^- \go \mu^+ \mu^-$ the deviation from the SM
amounts to $-4${\%} with the electron beam polarized in the right-handed mode by 80{\%}
$(P_{e^-} = 0.8)$ for $\theta_H \sim 0.09$, whereas there appears negligible deviation 
with the electron beam polarized in the left-handed mode by 80{\%} $(P_{e^-} = -0.8)$.
In the forward-backward asymmetry $A_{FB} (\mu^+ \mu^-)$
the deviation from the SM becomes $-2${\%} for $P_{e^-} = 0.8$.
These deviations can be seen at 250$\,$GeV ILC even 
with 250$\,$fb$^{-1}$ data.\cite{Bilokin2017, Fujii2017}
In the B-model the pattern of the polarization dependence is reversed.

So far quarks and leptons in the gauge-Higgs unification models have been 
incorporated  generation by generation so that the flavor mixing among 
quarks and leptons is left unexplained.
In this paper we tackle the flavor mixing in the quark sector.\cite{Adachi2010, Cacciapaglia2008}
We will argue in the B-model that the Cabibbo-Kobayashi-Maskawa (CKM) 
mixing matrix in the charged current interaction is reproduced with brane interactions.
These brane interactions generally lead to flavor-changing neutral current (FCNC) 
interactions.  It will be shown that the FCNC interactions are naturally suppressed 
in the gauge-Higgs unification as a consequence of the gauge invariance.    
The FCNC interaction is suppressed by a factor of $(m_b/m_\KK)^2 \sim 10^{-6}$ where 
$m_b$ and $m_\KK$ are the bottom quark mass and the KK mass scale.
It is also shown that induced flavor-changing Yukawa interactions are  extremely small.

We stress that the natural suppression of FCNC in the gauge-Higgs unification results from 
the gauge-invariance and the orbifold structure, without relying on additional symmetry
or mechanism.  We present rigorous treatment of deriving and evaluating the CKM matrix
and $Z$ couplings in the quark sector in the gauge-Higgs unification.  We also give 
simple explanation in the effective theory of quarks and relevant heavy fields to illuminate
the mechanism of suppressing FCNC interactions.

In section 2 the minimal GUT inspired $SU(3)_C \times SO(5)\times U(1)_X $ model of
gauge-Higgs unification is described with brane interactions.  
In section 3 mass spectra and wave functions of gauge bosons and quarks are
derived.  Detailed derivation of the mass spectrum and mixing in the down-type quark sector
is given.  In section 4 an effective theory in 4D is formulated for quarks and $SO(5)$ singlet
heavy fermion fields.  We show how mass terms connecting down quarks and singlet fields
lead to flavor mixing.  It also illuminates how FCNC interactions are naturally suppressed.
In section 5 we evaluate $W$ and $Z$ couplings of quarks, using the wave functions obtained 
in  section 3.  The gauge couplings turn out very close to those in the SM.  
It is confirmed that FCNC interactions are naturally suppressed.
Section 6 is devoted to summary and discussions.   Basis functions used in the text are 
summarized in the appendix.

\section{Model} 

The GUT inspired $SU(3)_C \times SO(5)\times U(1)_X (\equiv {\cal G})$ gauge-Higgs unification 
has been introduced  in ref.\ \cite{FHHOY2019a}. 
It is defined in the Randall-Sundrum (RS) warped space with metric  given by 
\begin{align}
ds^2= g_{MN} dx^M dx^N =
e^{-2\sigma(y)} \eta_{\mu\nu}dx^\mu dx^\nu+dy^2,
\label{Eq:5D-metric}
\end{align}
where $M,N=0,1,2,3,5$, $\mu,\nu=0,1,2,3$, $y=x^5$,
$\eta_{\mu\nu}=\mbox{diag}(-1,+1,+1,+1)$,
$\sigma(y)=\sigma(y+ 2L)=\sigma(-y)$,
and $\sigma(y)=ky$ for $0 \le y \le L$.
In terms of the conformal coordinate $z=e^{ky}$
($1\leq z\leq z_L=e^{kL}$) in the region $0 \leq y \leq L$ 
\begin{align}
ds^2=  \frac{1}{z^2} \bigg(\eta_{\mu\nu}dx^{\mu} dx^{\nu} + \frac{dz^2}{k^2}\bigg) .
\label{Eq:5D-metric-2}
\end{align}
The bulk region $0<y<L$ ($1<z<z_L$) is anti-de Sitter (AdS) spacetime 
with a cosmological constant $\Lambda=-6k^2$, which is sandwiched by the
UV brane at $y=0$ ($z=1$) and the IR brane at $y=L$ ($z=z_L$).  
The KK mass scale is $m_{\rm KK}=\pi k/(z_L-1) \simeq \pi kz_L^{-1}$
for $z_L\gg 1$.

Let us denote gauge fields of $SU(3)_C$,  $SO(5)$, and $U(1)_X$ by
$A_M^{SU(3)_C}$, $A_M^{SO(5)}$, and $A_M^{U(1)_X}$, respectively.
The orbifold boundary conditions (BC) are given by
\begin{align}
&\begin{pmatrix} A_\mu \cr  A_{y} \end{pmatrix} (x,y_j-y) =
P_{j} \begin{pmatrix} A_\mu \cr  - A_{y} \end{pmatrix} (x,y_j+y)P_{j}^{-1}
\label{Eq:BC-gauge}
\end{align}
for each gauge field where $(y_0, y_1) = (0, L)$.  In terms of  
\begin{align}
P_{\bf 3}^{SU(3)}&=I_3,\nonumber\\
P_{\bf 4}^{SO(5)}&=\mbox{diag}\left(I_{2},-I_{2}\right),\nonumber\\
P_{\bf 5}^{SO(5)}&=\mbox{diag}\left(I_{4},-I_{1}\right) , 
\label{Eq:SO5-BCs}
\end{align}
$P_0=P_1= P_{\bf 3}^{SU(3)}$ for  $A_M^{SU(3)_C}$  and 
$P_0=P_1= 1$ for $A_M^{U(1)_X}$.  
$P_0=P_1 = P_{\bf 5}^{SO(5)}$ for $A_M^{SO(5)}$ in the vector
representation and $P_{\bf 4}^{SO(5)}$ in the spinor representation, respectively.
$P_{\bf 4}^{SO(5)}$ and $P_{\bf 5}^{SO(5)}$ break $SO(5)$ to 
$SO(4) \simeq SU(2)_L \times SU(2)_R$.
$W$, $Z$ bosons and $\gamma$ (photon) are zero modes  in the $SO(4)$ part
of $A_\mu^{SO(5)}$, whereas the 4D Higgs boson is a zero mode in the 
$SO(5)/SO(4)$ part of $A_y^{SO(5)}$.

Matter fields are introduced both in 5D bulk and on the UV brane.
They are listed in Table~\ref{Tab:matter}.
Quark multiplets $({\bf 3}, {\bf 4})_{\frac{1}{6}}$ and $({\bf 3}, {\bf 1})_{-\frac{1}{3}}^\pm$
are introduced in the 5D bulk in three generations.  
They are denoted as
$\Psi_{({\bf 3,4})}^{\alpha}(x,y)$ and $\Psi_{({\bf 3,1})}^{\pm \alpha}(x,y)$
$(\alpha=1,2,3)$.  
$\Psi_{({\bf 3,4})}^{\alpha}$ and $\Psi_{({\bf 3,1})}^{\pm \alpha}$ intertwine with each other.  
These fields obey boundary conditions 
\begin{align}
&\Psi_{({\bf 3,4})}^{\alpha} (x, y_j - y) = 
- P_{\bf 4}^{SO(5)} \gamma^5 \Psi_{({\bf 3,4})}^{\alpha} (x, y_j + y) ~, \cr
&\Psi_{({\bf 3,1})}^{\pm \alpha}  (x, y_j - y) =
\mp \gamma^5 \Psi_{({\bf 3,1})}^{\pm \alpha}  (x, y_j + y) ~.
\label{quarkBC1}
\end{align}
With (\ref{quarkBC1}) the parity of quark fields are summarized in Table \ref{Tab:parity}
with names adopted in the present paper.

\begin{table}[tbh]
{
\renewcommand{\arraystretch}{1.2}
\begin{center}
\caption{${\cal G} = SU(3)_C\times SO(5) \times U(1)_X$ content of matter fields  is shown
in the  GUT inspired model (B model) and previous model (A model).
In the  A model only $SU(3)_C\times SO(4) \times U(1)_X$
symmetry is preserved on the UV brane so that the $SU(2)_L \times SU(2)_R$ content
is shown for brane fields. The B model is analyzed in the present paper.}
\vskip 10pt
\begin{tabular}{|c|c|c|}
\hline
&{B model} &{A model}\\
\hline \hline
quark
&$({\bf 3}, {\bf 4})_{\frac{1}{6}} ~ ({\bf 3}, {\bf 1})_{-\frac{1}{3}}^+ 
    ~ ({\bf 3}, {\bf 1})_{-\frac{1}{3}}^-$
&$({\bf 3}, {\bf 5})_{\frac{2}{3}} ~ ({\bf 3}, {\bf 5})_{-\frac{1}{3}}$ \\
lepton
&$\strut ({\bf 1}, {\bf 4})_{-\frac{1}{2}}$ 
&$({\bf 1}, {\bf 5})_{0} ~ ({\bf 1}, {\bf 5})_{-1}$  \\
\hline
dark fermion & $({\bf 3}, {\bf 4})_{\frac{1}{6}} ~ ({\bf 1}, {\bf 5})_{0}^+ ~ ({\bf 1}, {\bf 5})_{0}^-$ 
&$({\bf 1}, {\bf 4})_{\frac{1}{2}}$ \\
\hline \hline
brane fermion &$({\bf 1}, {\bf 1})_{0} $ 
&$\begin{matrix} ({\bf 3}, [{\bf 2,1}])_{\frac{7}{6}, \frac{1}{6}, -\frac{5}{6}} \cr
({\bf 1}, [{\bf 2,1}])_{\frac{1}{2}, -\frac{1}{2}, -\frac{3}{2}} \end{matrix}$\\
\hline
brane scalar &$({\bf 1}, {\bf 4})_{\frac{1}{2}} $ 
&$({\bf 1}, [{\bf 1,2}])_{\frac{1}{2}}$ \\
\hline
$\begin{matrix} {\rm symmetry ~of} \cr {\rm brane ~interactions} \end{matrix}$
&$SU(3)_C \times SO(5) \times U(1)_X$ &$SU(3)_C \times SO(4) \times U(1)_X$ \\
\hline
\end{tabular}
\label{Tab:matter}
\end{center}
}
\end{table}

\begin{table}[tbh]
\renewcommand{\arraystretch}{1.2}
\begin{center}
\caption{Parity assignment $(P_0, P_1)$ of quark multiplets in the bulk.
In the third column $G_{22}=SU(2)_L\times SU(2)_R$ content is shown.
Brane scalar field $\Phi_{({\bf 1}, {\bf 4})}$ is also  listed at the bottom for convenience.
}
\vskip 10pt
\begin{tabular}{|c|c|c|c|c|c|}
\hline
field & ${\cal G}$ & $G_{22}$ &left-handed &right-handed &name\\
\hline
$\Psi_{({\bf 3,4})}^{\alpha}$ &$({\bf 3,4})_{\frac{1}{6}}$ &$\, [{\bf 2} , {\bf  1}] \,$
&$(+,+)$ &$(-,-)$ &$\begin{matrix} u & c & t \cr d & s & b\end{matrix}$\\
\cline{3-6}
&&$[{\bf 1} , {\bf  2}]$ 
&$(-,-)$ &$(+,+)$ &$\begin{matrix} u'  & c' & t' \cr d' & s' & b' \end{matrix}$\\
\hline
$\Psi_{({\bf 3,1})}^{\pm \alpha}$ &$({\bf 3,1})_{-\frac{1}{3}}$ 
&$[{\bf 1} , {\bf  1}]$
&$(\pm ,\pm )$ &$(\mp , \mp )$ &$D^{\pm}_d ~ D^{\pm}_s ~ D^{\pm}_b$\\
\hline \hline
$\Phi_{({\bf 1}, {\bf 4})}$ &$({\bf 1,4})_{\frac{1}{2}}$ 
&$[{\bf 2} , {\bf  1}]$
&$\cdots$ & $\cdots$ &$\Phi_{[{\bf 2}, {\bf 1}]}$\\
\cline{3-6}
&&$[{\bf 1} , {\bf 2}]$ & $\cdots$ & $\cdots$ &$\Phi_{[{\bf 1}, {\bf 2}]}$\\
\hline
\end{tabular}
\label{Tab:parity}
\end{center}
\end{table}

The action of each gauge field, $A_M^{SU(3)_C}$, $A_M^{SO(5)}$, or $A_M^{U(1)_X}$,
is given by 
\begin{align}
S_{\rm bulk}^{\rm gauge}&=
\int d^5x \sqrt{-\det G} \, \bigg[ -\tr \bigg( \frac{1}{4} F^{MN} F_{MN}
+ \frac{1}{2\xi}(f_{\rm gf})^2 + {\cal L}_{\rm gh} \bigg) \bigg],
\label{Eq:Action-bulk-gauge}
\end{align}
where $\sqrt{-\det G}=1/k z^5$.
Field strengths  are defined by  $F_{MN} = \dd_M A_N - \dd_N A_M -i  g [A_M,A_N]$
with each 5D gauge coupling constant $g$.  
The gauge fixing and ghost terms are taken as
\begin{align}
f_{\rm gf}&=
 z^2 \bigg\{ \eta^{\mu\nu} {\cal D}^{\rm c}_\mu A^{\rm q}_\nu
 +\xi k^2 z{\cal D}^{\rm c}_z \Big( \frac{1}{z} A^{\rm q}_z\Big) \bigg\}, \cr
\noalign{\kern 10pt} 
{\cal L}_{\rm gh}& =
\bar{c} \, \Big\{ \eta^{\mu\nu} {\cal D}_\mu^{\rm c} {\cal D}_\nu
+ \xi k^2 z {\cal D}_z^{\rm c} \frac{1}{z} {\cal D}_z \Big\} \,  c ~,
\end{align}
where  $A_M=A_M^c+A_M^q$.
${\cal D}_M^{c}B=\partial_M B-ig[A_M^c,B]$ and
${\cal D}_M^{c+q}B=\partial_M B-ig[A_M,B]$ 
where $B=A_\mu^q$, ${A_z^q/z}$ and $c$.
Only $A_z$ component of $A_M^{SO(5)}$ has non-vanishing classical background $A_z^c$.

Each fermion multiplet $\Psi (x,y)$ in the bulk has its own bulk-mass parameter $c$.\cite{Gherghetta2000}
The covariant derivative is given by
\begin{align}
&{\cal D}(c)= \gamma^A {e_A}^M
\bigg( D_M+\frac{1}{8}\omega_{MBC}[\gamma^B,\gamma^C]  \bigg) -c\sigma'(y) ~, \cr
\noalign{\kern 5pt}
&D_M =  \dd_M - ig_S A_M^{SU(3)} -i g_A A_M^{SO(5)}  -i g_B Q_X A_M ^{U(1)} ~. 
\label{covariantD}
\end{align}
Here $\sigma' = d\sigma(y)/dy$ and $\sigma'(y) =k$ for $0< y < L$. 
$g_S$, $g_A$, $g_B$ are $SU(3)_C$, $SO(5)$, $U(1)_X$ gauge coupling constants.
The bulk part of the action for the quark multiplets are given by
\begin{align}
S_{\rm bulk}^{\rm quark} &=  \int d^5x\sqrt{-\det G} \,
\sum_{\alpha = 1}^3  \bigg\{ \overline{\Psi}{}_{\bf (3,4)}^\alpha   {\cal D} (c_\alpha) \Psi_{\bf (3,4)}^\alpha
+ \overline{\Psi}{}_{\bf (3,1)}^{+ \alpha}   {\cal D} (c_{D_\alpha^+} ) \Psi_{\bf (3,1)}^{+ \alpha} \cr
\noalign{\kern 5pt}
&\hskip 0.5cm
+ \overline{\Psi}{}_{\bf (3,1)}^{- \alpha}   {\cal D} (c_{D_\alpha^-} ) \Psi_{\bf (3,1)}^{- \alpha} 
-  m_{D_\alpha}  \Big( \overline{\Psi}{}_{\bf (3,1)}^{+ \alpha} \Psi_{\bf (3,1)}^{- \alpha} 
+ \overline{\Psi}{}_{\bf (3,1)}^{- \alpha} \Psi_{\bf (3,1)}^{+ \alpha} \Big) \bigg\} ,
\label{fermionAction1}
\end{align} 
where $\overline{\Psi} = i \Psi^\dagger \gamma^0$.
The bulk mass parameters of the $SO(5)$ spinor multiplets are denoted as
$(c_1, c_2, c_3) = (c_u, c_c, c_t)$ below as each $c_\alpha$ is determined from the mass of each 
up-type quark.  For the $SO(5)$ singlet multiplets we consider the case 
$c_{D_\alpha^+} = c_{D_\alpha^-} \equiv c_{D_\alpha}$ in the present paper.
(An alternative choice $c_{D_\alpha^+} = - c_{D_\alpha^-} $ is also possible. See ref.\ \cite{FHHOY2019a}.)

The action for the brane scalar field $\Phi_{({\bf 1}, {\bf 4})}(x)$ is given by
\begin{align}
&S_{\rm brane}^{\Phi}  = 
\int d^5x\sqrt{-\det G} \,  \delta(y) \cr
\noalign{\kern 5pt}
&\hskip 1.5cm
\times \Big\{ 
-(D_\mu\Phi_{({\bf 1,4})})^{\dag}D^\mu\Phi_{({\bf 1,4})}
-\lambda_{\Phi_{({\bf 1,4})}}
\big(\Phi_{({\bf 1,4})}^\dag\Phi_{({\bf 1,4})} - |w|^2  \big)^2 \Big\} , \cr
\noalign{\kern 5pt}
&D_\mu\Phi_{({\bf 1,4})}=
\bigg\{\partial_\mu- ig_A   \sum_{\alpha=1}^{10}   A_{\mu}^{\alpha} T^{\alpha}  
 -ig_B Q_X B_\mu  \bigg\}\Phi_{({\bf 1,4})} ~.
\label{Action-branescalar}
\end{align}
A spinor {\bf 4} of $SO(5)$ is decomposed to
$[{\bf 2}, {\bf 1}] \oplus [{\bf 1}, {\bf 2}]$ of $SO(4) \simeq SU(2)_L \times SU(2)_R$.
$\Phi_{({\bf 1}, {\bf 4})}$ develops a nonvanishing vacuum expectation value (VEV);
\begin{align}
\Phi_{({\bf 1,4})} =
\begin{pmatrix} \Phi_{[{\bf 2,1}]} \cr \Phi_{[{\bf 1,2}]} \end{pmatrix} , ~~
\la  \Phi_{[{\bf 1,2}]} \ra = \begin{pmatrix} 0 \cr w \end{pmatrix} ,
\label{scalarVEV}
\end{align}
which reduces the symmetry ${\cal G}' = SU(3)_C\times SO(4) \times U(1)_X$ to 
the SM gauge group  $G_{\rm SM}=SU(3)_C\times SU(2)_L\times U(1)_Y$.
It is assumed that $w \gg m_\KK$,  which ensures that  boundary conditions
for the 4D components of gauge fields corresponding to broken generators in the breaking
$SU(2)_R \times U(1)_X \go  U(1)_Y$ obey effectively Dirichlet conditions at the UV brane
for low-lying KK modes.\cite{Furui2016}
Accordingly the mass of the neutral physical mode of $\Phi_{({\bf 1,4})}$ is much larger 
than $m_\KK$.

There are brane interactions on the UV brane which are invariant under 
${\cal G} = SU(3)_C\times SO(5) \times U(1)_X$.
\begin{align}
&
 S_{\rm brane}^{\rm int}=
- \int d^5x\sqrt{-\det G} \, \delta(y) \, 
\bigg\{ \sum_{\alpha, \beta} \kappa_{\alpha\beta} \,
\overline{\Psi}{}_{({\bf 3,4})}^{\alpha} \Phi_{({\bf 1,4})}
 \Psi_{({\bf 3,1})}^{+\beta}  + {\rm h.c.} \bigg\}  ~, 
\label{BraneInt1}
\end{align}
where $\kappa_{\alpha\beta}$'s are coupling constants.
If only  the gauge invariance under  ${\cal G}' $ were imposed,   
there would appear  additional brane interactions.  
Instead of (\ref{BraneInt1}) one would have
\begin{align}
\sum_{\alpha, \beta} \Big\{ \kappa_{\alpha\beta}^{(1)} \,
\overline{\Psi}{}_{({\bf 3, [2,1]})}^{\alpha} \Phi_{({\bf 1, [2,1]})}
 \Psi_{({\bf 3,1})}^{+\beta}  
 + \kappa_{\alpha\beta}^{(2)} \,
\overline{\Psi}{}_{({\bf 3,[1,2]})}^{\alpha} \Phi_{({\bf 1,[1,2]})}
 \Psi_{({\bf 3,1})}^{+\beta}  \Big\}
+ {\rm h.c.}
\label{BraneInt2}
\end{align}
in the Lagrangian density.   The invariance under ${\cal G}$ implies 
$\kappa_{\alpha\beta}^{(1)}  = \kappa_{\alpha\beta}^{(2)} $.
For fermion fields we define $\check \Psi = z^{-2} \Psi$.
With nonvanishing VEV $\la \Phi_{({\bf 1},{\bf 4})} \ra \not= 0 $,  (\ref{BraneInt1})
generates mass terms
\begin{align}
&S_{\rm brane\ mass}^{\rm fermion}=
\int d^5x\sqrt{-\det G} \, \delta(y)
 \Big\{  \sum_{\alpha, \beta} 2\mu_{\alpha\beta} \, 
\overline{\check{d}\,}{}_{R}^{\prime\alpha} \check{D}_{L}^{+\beta}  +\mbox{h.c.} \Big\} ~,
\label{braneFmass1}
\end{align}
where $2\mu_{\alpha\beta}=\sqrt{2} \, \kappa_{\alpha\beta} \, w$, 
$(d^{\prime 1}, d^{\prime 2}, d^{\prime 3}) = (d', s', b')$ and 
$(D^{+ 1}, D^{+ 2}, D^{+ 3}) = (D_d^+, D_s^+, D_b^+)$. 
Only the $\kappa_{\alpha\beta}^{(2)}$ part in the decomposition (\ref{BraneInt2})
gives rise to mass terms.
Brane interaction of the form 
$\overline{\Psi}{}_{({\bf 3,4})}^{\alpha} \Phi_{({\bf 1,4})} \Psi_{({\bf 3,1})}^{-\beta}$
is possible, which, however,  does not yield a mass term as $D_L^{- \beta} |_{y=0} = 0$ 
due to the BC.
It will be shown below that the brane interactions (\ref{BraneInt1}) lead to the flavor mixing,
yielding the CKM matrix in the charged current interactions.  
We stress that the brane interactions (\ref{BraneInt1})  respect  full 
${\cal G} =SU(3)_C\times SO(5) \times U(1)_X$ gauge invariance.
It may be contrasted to the earlier attempts \cite{Adachi2010, Cacciapaglia2008} 
to incorporate flavor mixing  in higher dimensional theories where only 
$SU(3)_C\times SU(2)_L \times U(1)_Y$ gauge invariance is respected.
We note the same mass terms are generated from (\ref{BraneInt2}) so that
the results obtained below remain valid even with only the ${\cal G}'$ invariance
imposed on the brane so long as
$|\kappa_{\alpha\beta}^{(1)}/\kappa_{\alpha\beta}^{(2)}|$ is not extremely large.

Nonvanishing  VEV $\la \Phi_{({\bf 1},{\bf 4})} \ra$ also breaks $SU(2)_R \times U(1)_X$ to $U(1)_Y$.
$U(1)_Y$ gauge field $B_M^Y$ is given in terms of $SU(2)_R$ gauge fields $A_M^{a_R}$ and 
$U(1)_X$ gauge field $B_M$  by
\begin{align}
&B_M^Y = s_\phi A_M^{3_R} + c_\phi  B_M ~, \cr
\noalign{\kern 5pt}
&c_\phi  = \frac{g_A}{\sqrt{g_A^2+g_B^2}} ~, ~~
s_\phi  = \frac{g_B}{\sqrt{g_A^2+g_B^2}} ~,
\label{U1Y}
\end{align}
where $g_A$ and $g_B$ are gauge couplings in $SO(5)$ and $U(1)_X$, respectively.
The 5D $U(1)_{Y}$ gauge coupling   is given by $ g_Y^{\rm 5D} =g_A s_\phi$.
The 4D $SU(2)_L$ gauge coupling is given by $g_w = g_A/\sqrt{L}$.  

The 4D Higgs boson doublet $\phi_H(x)$ is the zero mode contained in the 
$A_z = (kz)^{-1} A_y$ component;
\begin{align}
A_z^{(j5)} (x, z) &= \frac{1}{\sqrt{k}} \, \phi_j (x) u_H (z) + \cdots , ~~
u_H (z) = \sqrt{ \frac{2}{z_L^2 -1} } \, z ~, \cr
\noalign{\kern 5pt}
\phi_H(x) &= \frac{1}{\sqrt{2}} \begin{pmatrix} \phi_2 + i \phi_1 \cr \phi_4 - i\phi_3 \end{pmatrix} .
\label{4dHiggs}
\end{align}
Without loss of generality we assume $\la \phi_1 \ra , \la \phi_2 \ra , \la \phi_3 \ra  =0$ and  
$\la \phi_4 \ra \not= 0$, 
which is related to the Aharonov-Bohm (AB) phase $\theta_H$ in the fifth dimension by
$\la \phi_4 \ra  = \theta_H f_H$ where
\begin{align}
&f_H  = \frac{2}{g_w} \sqrt{ \frac{k}{L(z_L^2 -1)}} ~.
\label{fH1}
\end{align}

\section{Mass spectrum and wave functions} 

Manipulations are simplified in the twisted gauge \cite{Falkowski2007, HS2007}
defined by  an $SO(5)$ gauge transformation
\begin{align}
&\tilde A_M (x,z) = \Omega A_M \Omega^{-1} 
+ \frac{i}{g_A} \, \Omega \,\dd_M \Omega^{-1} ~, \cr
\noalign{\kern 5pt}
& \Omega (z)  = \exp \Big\{ i \theta (z) T^{(45)} \Big\}  ~,~~
\theta (z) = \theta_H \, \frac{z_L^2 - z^2}{z_L^2 - 1} ~, 
\label{twisted1}
\end{align}
where $T^{jk}$'s are $SO(5)$ generators and  
$A_M = 2^{-1/2} \sum_{1 \le j<k \le 5} A_M^{(jk)} T^{jk}$.
In the twisted gauge the background field vanishes ($\tilde \theta_H = 0$) so that
all fields satisfy free equations in the RS space  in the bulk.  Boundary conditions
at the UV brane are modified,  whereas boundary conditions at the IR brane remain 
the same as in the original gauge.

\subsection{Gauge fields}

The masses of $W$ and $Z$ bosons at the tree level, $m_W = k \lambda_W$ and 
$m_Z= k \lambda_Z$,  are determined by 
\begin{align}
&2S(1;\lambda_W) C'(1;\lambda_W)+\lambda_W \sin^2\theta_H =  0 ~, \cr
\noalign{\kern 5pt}
&2S(1;\lambda_Z) C'(1;\lambda_Z)+(1+s_\phi^2)\lambda_Z\sin^2\theta_H =  0 ~, 
\label{WZmass1}
\end{align}
where functions $S(z;\lambda)$ and   $C(z;\lambda)$ are given in (\ref{gaugeF1})
and $s_\phi$ is defined in (\ref{U1Y}).
The masses are approximately given by
\begin{align}
m_W &\simeq \sqrt{\frac{k}{L}} \, z_L^{-1}\sin\theta_H 
\simeq\frac{\sin\theta_H}{\pi\sqrt{kL}} \, m_{\rm KK} ~, \cr
\noalign{\kern 5pt}
m_Z &\simeq \sqrt{1 + s_\phi^2} \, m_W ~.
\label{WZmass2}
\end{align}
$s_\phi$ is related to the Weinberg angle  at the tree level by 
$\sin \theta_W^0 = s_\phi/ \sqrt{1 + s_\phi^2}$.

Let us define 
\begin{align}
A_M^{a_L} &= \frac{1}{\sqrt{2}} 
\Big( \frac{1}{2} \eps^{abc} A_M^{(bc)} + A_M^{(a4)} \Big) ~, \cr
\noalign{\kern 5pt}
A_M^{a_R} &= \frac{1}{\sqrt{2}} 
\Big( \frac{1}{2} \eps^{abc} A_M^{(bc)} - A_M^{(a4)} \Big) ~, \cr
\noalign{\kern 5pt}
A_M^{\hat p} &= A_M^{(p5)} ~,
\label{gaugeA1}
\end{align}
where $a,b,c= 1 \sim 3$ and $p = 1 \sim 4$.
$A_M^{a_L}$ and $A_M^{a_R}$ are gauge fields of $SU(2)_L$ and $SU(2)_R$.  
For $W$ and $Z$ bosons and photon $\gamma$ we define
\begin{align}
\left[ \begin{matrix}  \mathring{W}_\mu (x,z) \cr   \mynoalign
\mathring{W}_\mu^S (x,z) \end{matrix} \right]
&=\sqrt{k} \, W_\mu (x) \, \frac{1}{\sqrt{r_W}} 
\left[ \begin{matrix} C(z, \lambda_W) \cr \mynoalign
\hat S(z, \lambda_W) \end{matrix} \right] , \cr
\noalign{\kern 7pt}
\left[ \begin{matrix}  \mathring{Z}_\mu (x,z) \cr  \mynoalign
\mathring{Z}_\mu^S (x,z) \end{matrix} \right]
&=\sqrt{k} \, Z_\mu (x) \, \frac{1}{\sqrt{r_Z}} 
\left[ \begin{matrix} C(z, \lambda_Z) \cr \mynoalign
\hat S(z, \lambda_Z) \end{matrix} \right] , \cr
\noalign{\kern 5pt}
&\hskip -1.5cm
\mathring{A}_\mu^\gamma 
= \sqrt{k} \, A_\mu^\gamma (x)  \, \frac{1}{\sqrt{k L}}  ~, ~~
\hat S(z, \lambda) = \frac{C(1, \lambda)}{S(1, \lambda)} \, S(z, \lambda) ~, 
\label{tower1}
\end{align}
where 
\begin{align}
&r_W = \int_1^{z_L} \frac{dz}{z} \big\{ (1+ c_H^2) \, C(z, \lambda_W)^2 
+ s_H^2 \hat S(z, \lambda_W)^2 \big\} ~, \cr
\noalign{\kern 5pt}
&r_Z = \int_1^{z_L} \frac{dz}{z} \Big\{ [ c_\phi^2 +(1+ s_\phi^2) c_H^2] \, C(z, \lambda_Z)^2 
+ (1+ s_\phi^2) s_H^2 \hat S(z, \lambda_Z)^2 \Big\} ~, \cr
\noalign{\kern 5pt}
&c_H = \cos \theta_H ~, ~~ s_H = \sin \theta_H ~.
\end{align}
Here $W_\mu(x)$, $Z_\mu (x)$ and $A_\mu^\gamma (x)$ represent canonically
normalized $W$, $Z$, and $\gamma$ fields, respectively.
Note that $\lambda_W z_L, \lambda_Z z_L \ll 1$.
For $\lambda z_L \ll 1$, $C(z, \lambda) \sim z_L$ and 
$\hat S (z. \lambda) \sim z_L (1 - z^2/z_L^2)$.

Couplings of $W$, $Z$, and $\gamma$ are obtained by inserting
\begin{align}
&\left[ \begin{matrix} \tilde A_\mu^{1_L} - i \tilde A_\mu^{2_L} \cr
\mynoalign
\tilde A_\mu^{1_R} - i \tilde A_\mu^{2_R} \cr
\mynoalign
\tilde A_\mu^{\hat 1} - i \tilde A_\mu^{\hat 2} \end{matrix} \right]=
\left[ \begin{matrix} (1 + c_H)  \mathring{W}_\mu \cr  \mynoalign
(1 - c_H)  \mathring{W}_\mu \cr  \mynoalign
- \sqrt{2} \, s_H \mathring{W}_\mu^S \end{matrix} \right], \cr
\noalign{\kern 7pt}
&\left[ \begin{matrix} \tilde A_\mu^{3_L} \cr \mynoalign
\tilde A_\mu^{3_R} \cr \mynoalign \tilde A_\mu^{\hat 3} \cr \mynoalign B_\mu
\end{matrix} \right]= \frac{\sqrt{\smash[b]{1+ s_\phi^2}}}{\sqrt{2}}
\left[ \begin{matrix} (1 + c_H)  \mathring{Z}_\mu \cr  \mynoalign
(1 - c_H)  \mathring{Z}_\mu \cr  \mynoalign
- \sqrt{2} \, s_H \mathring{Z}_\mu^S \cr  \mynoalign 0 \end{matrix} \right]
+ \frac{1}{\sqrt{\smash[b]{1 + s_\phi^2}}} \left[ \begin{matrix} 
s_\phi \cr \mynoalign  s_\phi \cr \mynoalign 0   \cr \mynoalign c_\phi \end{matrix} \right]
(\mathring{A}_\mu^\gamma - \sqrt{2} s_\phi \mathring{Z}_\mu )
\label{waveGauge1}
\end{align}
in the  $SO(5)$ gauge fields $\tilde A_\mu$ in the twisted gauge
and $U(1)_Y$ gauge field $B_\mu^Y$ in the action.

\subsection{Up-type quarks}

Up, charm, and top quarks are zero modes contained solely in 
the fields $\Psi_{({\bf 3,4})}^{\alpha}$ and there arises no mixing in generation.
The mass spectrum $m_q = k \lambda_q$  ($q= u, c, t$) is determined by
\begin{align}
&S_L (1;  \lambda, c_q)  S_R (1;  \lambda, c_q) +\sin^2\onehalf \theta_H =0 ~.
\label{Up-quark-mass1}
\end{align}
Basis functions for fermions, $S_{L/R} (z, \lambda, c)$ and $C_{L/R} (z, \lambda, c)$,
are given by (\ref{fermionF1}).
For the first and second generation $|c_u|, |c_c| > \onehalf$, whereas for the third
generation   $|c_t| < \onehalf$.  The masses are approximately given by 
\begin{align}
m_{u, c} &\sim  \pi^{-1}\sqrt{4c_{u, c}^2-1} \, z_L^{-|c_{u,c}|+0.5} 
\sin \onehalf \theta_H  \, m_{\rm KK} ~,  \cr
\noalign{\kern 5pt}
m_t &\sim  \pi^{-1}\sqrt{1-4c_{t}^2} \, \sin \onehalf \theta_H  \, m_{\rm KK} ~.
\label{Up-quark-mass2}
\end{align}

4D  fields denoted by $\hat u (x)$ appear in the $(u, u')$ components in 
the 5D fields $\Psi_{({\bf 3,4})}^{\alpha=1} (x,z)$. (See Table \ref{Tab:parity}.)
In the twisted gauge, 
\begin{align}
\left[ \begin{matrix} \tilde{\check u} \cr \mynoalign \tilde{\check u}{}' \end{matrix} \right]
&= \frac{\sqrt{k}}{\sqrt{r_u}} \, \bigg\{  \hat u_L(x) 
\left[ \begin{matrix} \bar c_H C_L (z;  \lambda_u, c_u) \cr \mynoalign
i \bar s_H \hat S_L (z;  \lambda_u, c_u) \end{matrix} \right] 
+ \hat u_R(x) \left[ \begin{matrix} \bar c_H S_R (z;  \lambda_u, c_u)  \cr \mynoalign
i \bar s_H \hat C_R (z;  \lambda_u, c_u) \end{matrix} \right] \bigg\} ~,  \cr
\noalign{\kern 5pt}
r_u &= \int_1^{z_L} dz \Big\{ \bar c_H^2 C_L (z; \lambda_u, c_u)^2
+ \bar s_H^2 \hat S_L (z; \lambda_u, c_u)^2 \Big\} \cr 
\noalign{\kern 5pt}
&= \int_1^{z_L} dz \Big\{ \bar c_H^2 S_R (z;  \lambda_u, c_u)^2
+ \bar s_H^2 \hat C_R (z; \lambda_u, c_u)^2 \Big\} ~,  \cr 
\noalign{\kern 5pt}
\bar c_H &= \cos \onehalf \theta_H ~, ~~ \bar s_H = \sin \onehalf \theta_H ~, \cr
\noalign{\kern 5pt}
\hat S_L &(z;  \lambda, c) = \frac{C_L(1;  \lambda, c)}{S_L(1;  \lambda, c)} \, S_L (z;  \lambda, c) ~, ~
\hat C_R(z;  \lambda, c) = \frac{C_L(1;  \lambda, c)}{S_L(1; \lambda, c)} \, C_R (z;  \lambda, c) ~.
\label{waveUp1}
\end{align}
The equality of the two expressions for $r_u$ is confirmed with the aid of (\ref{Up-quark-mass1}).
Formulas for  charm and top quark fields are obtained by substitution $u \go c, t$.

\subsection{Down-type quarks}

Down, strange and bottom quarks are  contained  in 
$\Psi_{({\bf 3,4})}^{\alpha}$ and $\Psi_{({\bf 3,1})}^{\pm \alpha}$.
By the brane interactions (\ref{BraneInt1}) and (\ref{braneFmass1}) all three generations 
mix with each other.
In ref.\ \cite{FHHOY2019a}  the mass spectrum is determined in each generation
separately.  Generalization to the case with mixing is straightforward.  
We consider the case in which both $\Psi_{({\bf 3,1})}^{+ \alpha}$ and 
$\Psi_{({\bf 3,1})}^{- \alpha}$ have the same bulk mass parameters
$c_{D_\alpha^+} = c_{D_\alpha^-} \equiv c_{D_\alpha}$.
Without loss of generality we assume Dirac masses $m_{D_\alpha}$ in (\ref{fermionAction1})
are real.

For the sake of  clarity we adopt vector/matrix notation in the generation space.
Fermion fields are expressed in terms of  ``checked'' fields; $\check \psi = z^{-2} \psi$.
Write
\begin{align}
&\vec d = \begin{pmatrix} \check d \cr \check s \cr \check b \end{pmatrix}, ~
\vec d' = \begin{pmatrix} \check d' \cr \check s' \cr \check b' \end{pmatrix}, ~
\vec D^\pm = \begin{pmatrix} \check D_d^\pm \cr \check D_s^\pm \cr \check D_b^\pm \end{pmatrix}, \cr
\noalign{\kern 5pt}
&D_\pm^q = \begin{pmatrix} D_\pm (c_u) &&\cr &  D_\pm (c_c) & \cr &&  D_\pm (c_t) \end{pmatrix}, ~~
D_\pm (c) = \pm \frac{\dd}{\dd z} + \frac{c}{z} ~, \cr
\noalign{\kern 5pt}
&D_\pm^D = \begin{pmatrix} D_\pm (c_{D_d}) &&\cr &  D_\pm (c_{D_s}) & \cr &&  D_\pm (c_{D_b}) \end{pmatrix},  \cr
\noalign{\kern 5pt}
&\tilde m_D = \begin{pmatrix} \tilde m_{D_d} &&\cr &\tilde m_{D_s} &\cr &&\tilde m_{D_b} \end{pmatrix},
~~ \tilde m_{D_\alpha} = \frac{m_{D_\alpha}}{k} ~, \cr
\noalign{\kern 5pt}
&\mu =  \begin{pmatrix} \mu_{11} & \mu_{12} & \mu_{13} \cr
 \mu_{21} & \mu_{22} & \mu_{23} \cr  \mu_{31} & \mu_{32} & \mu_{33} \end{pmatrix} .
 \label{notation1}
\end{align}
In terms of two-component 4D Lorentz spinors ($\vec d_L , \vec d_R$ etc.) 
the equations of motion in the original gauge are given by
\begin{align}
\begin{matrix}  (a) : \cr \mysnoalign (b) : \end{matrix} ~~
 & \sigma^\mu \dd_\mu \begin{pmatrix}  \vec{d}_{L} \cr \vec{d}_{L}' \end{pmatrix}
-k\hat{D}_-^q \begin{pmatrix}  \vec{d}_{R} \cr \vec{d}_{R}' \end{pmatrix}  =0 ~, \cr
\noalign{\kern 5pt}
\begin{matrix}  (c) :\cr  \mysnoalign  (d) :\end{matrix} ~~
& \bar{\sigma}^\mu\partial_\mu 
 \begin{pmatrix}  \vec{d}_{R} \cr \check{d}_{R}' \end{pmatrix}
 -k\hat{D}_+^q \begin{pmatrix}  \vec{d}_{L} \cr \vec{d}_{L}' \end{pmatrix}
 = \delta(y)  \, 2\mu \begin{pmatrix} 0 \cr \vec{D}_{L}^{+} \end{pmatrix} , \cr
\noalign{\kern 5pt}
 (e): ~ 
 & \sigma^\mu \dd_\mu \vec{D}_L^+   -k D_-^D \vec{D}_{R}^+ 
  -\frac{k \tilde m_D }{z} \, \vec{D}_R^- =\delta(y) 2\mu^\dagger \vec{d}_R'  ~, \cr
\noalign{\kern 5pt}
(f): ~
 & \bar{\sigma}^\mu \dd_\mu \vec{D}_R^+ 
 - k D_+^D  \vec{D}_L^+ 
 - \frac{k \tilde m_D}{z} \, \vec{D}_L^-  = 0 ~,  \cr
 \noalign{\kern 5pt}
(g):~
 &   \sigma^\mu \dd_\mu \vec{D}_L^-  - k D _-^D \vec{D}_R^-
 -\frac{k \tilde m_D}{z}\, \vec{D}_R^+   = 0 ~, \cr
\noalign{\kern 5pt}
(h):~ 
 & \bar{\sigma}^\mu \dd_\mu \vec{D}_R^- 
 -k  D _+^D  \vec{D}_L^- -\frac{k \tilde m_D}{z} \, \vec{D}_L^+  =0 ~.
\label{downEq1}
\end{align}
The $\mu$ terms on the right side of the equations come from  the brane interaction
(\ref{braneFmass1}).  
The derivative $\hat D_\pm^q$ in Eqs.\ $(a)$-$(d)$ represents,
in each generation subspace,
\begin{align}
&\hat{D}_\pm (c) = D_\pm(c) \pm i \theta'(z) T^{45}  ~, \cr
\noalign{\kern 5pt}
&T^{45} = \onehalf \begin{pmatrix} 0&1 \cr 1 & 0 \end{pmatrix} ~~{\rm for}~
\begin{pmatrix} \check d \cr \check d' \end{pmatrix}, 
\begin{pmatrix} \check s\cr \check s' \end{pmatrix}, 
\begin{pmatrix} \check b \cr \check b' \end{pmatrix},
\label{derivativeD2}
\end{align}
where $\theta (z)$ is given by (\ref{twisted1}).
Note that the mass dimension of each coupling and field is
e.g., $[\check{d}_{R/L}]=2$, $[k]=1$ and $[\tilde m_D]=[\mu]=0$.

Boundary conditions at the IR brane ($z=z_L$) are, in the original gauge,  
\begin{align}
 \left\{
 \begin{array}{l}
  \vec d_{R}=0 ~, \cr \mysnoalign
  D_+^q  \vec d_L=0 ~, \cr \mysnoalign
  D_-^q \vec d_R' = 0 ~,\cr \mysnoalign
  \vec d_L'=0 ~,
 \end{array}
 \right. \quad 
 \left\{
 \begin{array}{l}
 \vec  D_R^+ = 0 ~, \cr \mysnoalign
  D_+^D  \vec D_L^+ = 0 ~,  \cr \mysnoalign
  D_-^D  \vec D_R^- = 0 ~, \cr \mysnoalign
 \vec  D_L^- = 0 ~. 
 \end{array}
 \right.
\label{downBC1}
\end{align}
Fields in the twisted gauge ($\tilde \chi$) are related to those in the original gauge  ($\chi$)  by
\begin{align}
&\chi = \begin{pmatrix} \cos \onehalf \theta (z) & -i \sin \onehalf \theta (z) \cr
\noalign{\kern 5pt}
-i \sin \onehalf \theta (z) & \cos \onehalf \theta (z)  \end{pmatrix}  \tilde{\chi} ~,  \cr
\noalign{\kern 5pt}
&\chi = \begin{pmatrix} \check d \cr \check d' \end{pmatrix}, ~
\begin{pmatrix} \check s \cr \check s' \end{pmatrix}, ~
\begin{pmatrix} \check b \cr \check b' \end{pmatrix} ,
\label{twisted2}
\end{align}
so that all fields in the twisted gauge obey the same boundary conditions as (\ref{downBC1}).

In the twisted gauge all fields in the bulk ($1 < z < z_L$) satisfy free equations with vanishing
background field $\tilde \theta_H =0$.   General solutions satisfying BC (\ref{downBC1}) are
\begin{align}
&\vec{\tilde d}_R = \begin{pmatrix}  \alpha_{d} S_R(z;\lambda, c_u) \cr 
\alpha_{s} S_R(z;\lambda, c_c) \cr \alpha_{b} S_R(z;\lambda, c_t) \end{pmatrix}, ~~
\vec{\tilde d}_L = \begin{pmatrix}  \alpha_{d} C_L(z;\lambda, c_u) \cr 
\alpha_{s} C_L(z;\lambda, c_c) \cr \alpha_{b} C_L(z;\lambda, c_t) \end{pmatrix},  \cr
\noalign{\kern 5pt}
&\vec{\tilde d}{}_R^{\, \prime} = \begin{pmatrix}  \alpha_{d'} C_R(z;\lambda, c_u) \cr 
\alpha_{s'} C_R(z;\lambda, c_c) \cr \alpha_{b'} C_R(z;\lambda, c_t) \end{pmatrix}, ~~
\vec{\tilde d}{}_L^{\, \prime}  = \begin{pmatrix}  \alpha_{d'} S_L(z;\lambda, c_u) \cr 
\alpha_{s'} S_L(z;\lambda, c_c) \cr \alpha_{b'} S_L(z;\lambda, c_t) \end{pmatrix},  \cr
\noalign{\kern 5pt}
&\vec{D}{}_R^+ = \begin{pmatrix}
a_{d}{\cal S}_{R2}(z;\lambda, c_{D_d}, \tilde m_{D_d})
+b_{d}{\cal S}_{R1}(z;\lambda, c_{D_d}, \tilde m_{D_d} ) \cr 
a_{s}{\cal S}_{R2}(z;\lambda, c_{D_s}, \tilde m_{D_s})
+b_{s}{\cal S}_{R1}(z;\lambda, c_{D_s}, \tilde m_{D_s} ) \cr 
a_{b}{\cal S}_{R2}(z;\lambda, c_{D_b}, \tilde m_{D_b})
+b_{b}{\cal S}_{R1}(z;\lambda, c_{D_b}, \tilde m_{D_b} ) 
\end{pmatrix} ,  \cr
\noalign{\kern 5pt}
&\vec{D}{}_L^+ = \begin{pmatrix}
a_{d}{\cal C}_{L2}(z;\lambda, c_{D_d}, \tilde m_{D_d})
+b_{d}{\cal C}_{L1}(z;\lambda, c_{D_d}, \tilde m_{D_d} ) \cr 
a_{s}{\cal C}_{L2}(z;\lambda, c_{D_s}, \tilde m_{D_s})
+b_{s}{\cal C}_{L1}(z;\lambda, c_{D_s}, \tilde m_{D_s} ) \cr 
a_{b}{\cal C}_{L2}(z;\lambda, c_{D_b}, \tilde m_{D_b})
+b_{b}{\cal C}_{L1}(z;\lambda, c_{D_b}, \tilde m_{D_b} ) 
\end{pmatrix} ,  \cr
\noalign{\kern 5pt}
&\vec{ D}{}_R^- = \begin{pmatrix}
a_{d}{\cal C}_{R1}(z;\lambda, c_{D_d}, \tilde m_{D_d})
+b_{d}{\cal C}_{R2}(z;\lambda, c_{D_d}, \tilde m_{D_d} ) \cr 
a_{s}{\cal C}_{R1}(z;\lambda, c_{D_s}, \tilde m_{D_s})
+b_{s}{\cal C}_{R2}(z;\lambda, c_{D_s}, \tilde m_{D_s} ) \cr 
a_{b}{\cal C}_{R1}(z;\lambda, c_{D_b}, \tilde m_{D_b})
+b_{b}{\cal C}_{R2}(z;\lambda, c_{D_b}, \tilde m_{D_b} ) 
\end{pmatrix} ,  \cr
\noalign{\kern 5pt}
&\vec{D}{}_L^- = \begin{pmatrix}
a_{d}{\cal S}_{L1}(z;\lambda, c_{D_d}, \tilde m_{D_d})
+b_{d}{\cal S}_{L2}(z;\lambda, c_{D_d}, \tilde m_{D_d} ) \cr 
a_{s}{\cal S}_{L1}(z;\lambda, c_{D_s}, \tilde m_{D_s})
+b_{s}{\cal S}_{L2}(z;\lambda, c_{D_s}, \tilde m_{D_s} ) \cr 
a_{b}{\cal S}_{L1}(z;\lambda, c_{D_b}, \tilde m_{D_b})
+b_{b}{\cal S}_{L2}(z;\lambda, c_{D_b}, \tilde m_{D_b} ) 
\end{pmatrix} .
\label{down-wave1} 
\end{align}
The tilde $\tilde{~}$ above each field indicates that it is in the twisted gauge.
Note $\vec{\tilde D}^\pm = \vec{D}^\pm$.
Functions $ {\cal S}_{R1}  (z; \lambda, c , \tilde m)$ etc.\ are defined in (\ref{MfermionBasis1}).  
The coefficients
\begin{align}
\vec \alpha = \begin{pmatrix}  \alpha_{d} \cr  \alpha_{s}  \cr  \alpha_{b} \end{pmatrix}, ~
\vec \alpha' = \begin{pmatrix}  \alpha_{d'} \cr  \alpha_{s'}  \cr  \alpha_{b'} \end{pmatrix}, ~
\vec a = \begin{pmatrix} a_{d} \cr  a_{s}  \cr  a_{b} \end{pmatrix}, ~
\vec b =  \begin{pmatrix} b_{d} \cr  b_{s}  \cr  b_{b} \end{pmatrix}
\label{coefficient1}
\end{align}
are determined such that BC at $z= 1^+$ ($y = +\ep$)  be satisfied.

To find BC at $z= 1^+$, first note that in the $y$ coordinate
\begin{align}
D_\pm (c) = \frac{e^{-\sigma(y)}}{k} \bigg\{ \pm \frac{\dd}{\dd y} + c \sigma' (y) \bigg\}.
\label{derivativeD3}
\end{align}
Fields $\vec d_L$, $\vec d'_R$, $D_L^+$ and $D_R^-$ are parity even at $y=0$, whereas
$\vec d_R$, $\vec d'_L$, $D_R^+$ and $D_L^-$ are parity odd.
We integrate the equations for parity odd fields, $(a)$, $(d)$, $(e)$ and $(h)$ in (\ref{downEq1}),
from $y=-\ep$ to $+\ep$ to find
\begin{align}
& \vec{d}_{R}(\epsilon)=0 ~, \cr
& \vec{d}'_{L}(\epsilon)  + \mu \vec{D}^+_{L}(0)=0 ~, \cr
&  \vec{D}^+_{R}(\epsilon) - \mu^\dagger \vec{d}'_{R}(0)=0 ~, \cr
& \vec{D}^-_{L}(\epsilon)=0 ~.
\label{downBC2}
\end{align}
For parity even fields we evaluate the equations $(b)$, $(c)$, $(f)$ and $(g)$ at $y=+\ep$, 
by using (\ref{downBC2}), to find
\begin{align}
& \hat{D}_-^q \vec{d}'_R +  \mu  \big\{  D_-^D \vec{D}^+_R +  \tilde{m}_D  \vec{D}^-_R \big\}  = 0 ~, \cr
& \hat{D}_+^q  \vec{d}_L=0 ~, \cr
&D_+^D  \vec{D}^+_L  - \mu^\dagger  D_+^D \vec{d}'_L    = 0~, \cr
&D_-^D \vec{D}^-_R + \tilde{m}_D  \vec{D}_R^+ =0 ~.
\label{downBC3}
\end{align}

Inserting (\ref{down-wave1}) into (\ref{downBC2}) and (\ref{downBC3}), one finds equations for 
the coefficient vectors in (\ref{coefficient1}).
The conditions (\ref{downBC2}) and (\ref{downBC3}) are split into two sets, one for
left-handed components and the other for right-handed components.  The two sets
yield equivalent conditions.  Making use of the relation (\ref{twisted2}) and equations 
$D_+ (C_L, S_L) = \lambda (S_R, C_R)$,  
$D_+ ( {\cal C}_{Lj},  {\cal S}_{Lj}) = \lambda ( {\cal S}_{Rj},  {\cal C}_{Rj}) -(\tilde m/z) ({\cal S}_{Lk},  {\cal C}_{Lk})$
[$(j,k) = (1,2), (2,1)$] etc., one finds for the set of left-handed components that
\begin{align}
(p_1):&~
\bar c_H S_R^q  \vec \alpha - i \bar s_H C_R^q \vec \alpha' = 0 ~, \cr
\noalign{\kern 5pt}
(p_2):&~
- i \bar s_H C_L^q  \vec \alpha + \bar c_H S_L^q  \vec \alpha'
+ \mu \big\{ {\cal C}_{L2}^D \vec a + {\cal C}_{L1}^D \vec b \big\} = 0 ~, \cr
\noalign{\kern 5pt}
(p_3):&~
 {\cal S}_{L1}^D \vec a +  {\cal S}_{L2}^D \vec b = 0 ~, \cr
\noalign{\kern 5pt}
(p_4):&~
 {\cal S}_{R2}^D \vec a +  {\cal S}_{R1}^D \vec b 
- \mu^\dagger \big\{ - i \bar s_H S_R^q \vec \alpha + \bar c_H C_R^q \vec \alpha' \big\} = 0 ~, 
\label{downBC4}
\end{align}
where
\begin{align}
&S_R^q = \begin{pmatrix} S_R  (1; \lambda , c_u) &&\cr & S_R (1;  \lambda , c_c) &\cr 
              &&S_R (1;  \lambda , c_t) \end{pmatrix},  \cr
\noalign{\kern 5pt}
&{\cal S}_{R j}^D = \begin{pmatrix} {\cal S}_{R j} (1; \lambda, c_{D_d} , \tilde m_{D_d} ) &&\cr 
& {\cal S}_{R j} (1; \lambda, c_{D_s} , \tilde m_{D_s} )  &\cr 
&&{\cal S}_{R j} (1; \lambda, c_{D_b} , \tilde m_{D_b} )  \end{pmatrix}, 
\label{downBC5}
\end{align}
and so on.  With the use of $(p_1)$ and $(p_3)$, $\vec \alpha'$ and $\vec b$ are expressed in 
terms of $\vec \alpha$ and $\vec a$, respectively.  Then $(p_2)$ and $(p_4)$ become
\begin{align}
& \frac{i}{\bar s_H} \big\{ \bar s_H^2 C_L^q + \bar c_H^2 S_L^q (C_R^q)^{-1} S_R^q \big\} \vec \alpha
- \mu \big\{ {\cal C}_{L2}^D - {\cal C}_{L1}^D ({\cal S}_{L2}^D )^{-1} {\cal S}_{L1}^D \big\} \vec a = 0 ~, \cr
\noalign{\kern 5pt}
&\big\{ {\cal S}_{R2}^D - {\cal S}_{R1}^D ({\cal S}_{L2}^D )^{-1} {\cal S}_{L1}^D \big\}  \vec a
+ \frac{i}{\bar s_H} \, \mu^\dagger S_R^q \vec \alpha = 0 ~.
\label{downBC6}
\end{align}
All matrices in (\ref{downBC6}) except for $\mu$ are diagonal.  Eliminating $\vec a$, one finds that
\begin{align}
&K(\lambda) \, S_R^q \, \vec \alpha = 0 ~, \cr
\noalign{\kern 5pt}
&K(\lambda) = \frac{S_L^q S_R^q + \bar s_H^2}{S_R^q C_R^q} +
\mu \, \frac{{\cal C}_{L1}^D  {\cal S}_{L1}^D - {\cal C}_{L2}^D  {\cal S}_{L2}^D}
                 {{\cal S}_{R1}^D  {\cal S}_{L1}^D - {\cal S}_{R2}^D  {\cal S}_{L2}^D} \,  \mu^\dagger ~.
\label{downmass1}
\end{align}
The mass spectrum $m_n = k \lambda_n$ of down-type quarks is obtained by
\begin{align}
\det \, K(\lambda_n) = 0 ~.
\label{downmass2}
\end{align}
Three lowest roots correspond to $m_d, m_s, m_b$.
In the $\mu \go 0$ limit, the down-quark spectrum is given by
$\det (S_L^q S_R^q + \bar s_H^2) =0$,  the same formula as for
the up-quark spectrum, and the spectrum of $D^\pm$ fields is given by 
$\det ({\cal S}_{R1}^D  {\cal S}_{L1}^D - {\cal S}_{R2}^D  {\cal S}_{L2}^D) = 0$.
As pointed out in ref.\ \cite{FHHOY2019a}, the spectrum  for $c_u, c_c >0$ contains  
exotic light fermions when $\mu \not= 0$.  For this reason we take $c_u, c_c, c_t < 0$.
We shall see below that gauge couplings of quarks remain very close to those in the SM
for $c_u, c_c, c_t < 0$ as well.

The coefficient vector $S_R^q \, \vec \alpha$ of each down-type quark  
is an eigenvector of $K(\lambda_n)$ with a zero eignevalue.
Once $\vec \alpha$ is determined, $\vec a$, and $\vec \alpha'$ and $\vec b$ are determined.  
Consequently the wave functions in (\ref{down-wave1})
are determined, with which all gauge couplings can be evaluated.

\section{Effective theory of CKM and FCNC}

Before evaluating the $W, Z$ gauge couplings of quarks by using exact wave functions
obtained in section 3, it is instructive to write down an effective theory of 
relevant fields to see how the brane interactions $\mu$ lead to flavor 
mixing and FCNC.  The effective theory illuminates also how FCNC interactions are 
naturally suppressed.

One crucial ingredient for lifting the degeneracy in the masses of  up and
down quarks is that right-handed component of down  quark is 
mixture of $d'$ and $D_d^\pm$.    As confirmed in the next section, dominant
part of physical down-type quarks,  $(\hat d_R, \hat s_R, \hat b_R)$, are contained in 
$(D_{dR}^-, D_{sR}^-, D_{bR}^-)$.   
It also  assures that the $W$ boson barely couples to right-handed components of 
physical up-type quarks 
as they  are contained solely in $\Psi_{({\bf 3}, {\bf 4})}^\alpha$.

\subsection{Mass matrix}

To simplify expressions, we use the following vector notation for 4D fermion fields in this section.
\begin{align}
&\vec u  = \begin{pmatrix} u \cr c \cr t \end{pmatrix}, ~ 
\vec u'  = \begin{pmatrix} u' \cr c' \cr t' \end{pmatrix}, ~ 
\vec d  = \begin{pmatrix} d \cr s \cr b \end{pmatrix}, ~ 
\vec d'  = \begin{pmatrix} d' \cr s' \cr b' \end{pmatrix}, ~ 
\vec D  = \begin{pmatrix} D_d \cr D_s \cr D_b \end{pmatrix}.
\label{4Dnotation1}
\end{align}
The masses of up-type quarks are generated solely by the Hosotani mechanism.
The effective mass terms in four dimensions are written as
\begin{align}
&{\cal L}_m^{\rm up} = - \Big\{   \vec{\bar u}_L{}^t \, M_{\rm up}  \, \vec{u}_R'  + {\rm h.c.}  \, \Big\} , \cr
\noalign{\kern 5pt}
&
M_{\rm up} = \begin{pmatrix} m_u &&\cr & m_c &\cr && m_t \end{pmatrix} .
\label{upEffMass1}
\end{align}
For down-type quarks the effective mass terms are written as
\begin{align}
&{\cal L}_m^{\rm down} = - \bigg\{  (  \vec{\bar d}_L{}^t,   \vec{\bar D}_L{}^t ) \, {\cal M}_{\rm down} 
\begin{pmatrix}   \vec{d}_R'  \cr  \vec{D}_R \end{pmatrix}  + {\rm h.c.} \, \bigg\} , \cr
\noalign{\kern 5pt}
&{\cal M}_{\rm down} = \begin{pmatrix} M_{\rm up} & 0 \cr \check \mu & \check m_D \end{pmatrix} , \cr
\noalign{\kern 5pt}
&\check \mu =  \begin{pmatrix} \check \mu_{11} & \check \mu_{12} & \check \mu_{13} \cr
\check  \mu_{21} & \check \mu_{22} &\check  \mu_{23} \cr  
\check \mu_{31} &\check  \mu_{32} &\check \mu_{33} \end{pmatrix} , ~~
\check m_D = \begin{pmatrix} \check m_{D_d} && \cr &\check m_{D_s} & \cr && \check m_{D_b} \end{pmatrix}.
\label{downEffMass1}
\end{align}
The Hosotani mechanism generates degenerate masses, the $M_{\rm up}$ term in ${\cal M}_{\rm down} $, 
for the components in $\Psi_{({\bf 3}, {\bf 4})}^\alpha$.
$D_{\alpha L}$ ($D_{\alpha R}$) is approximately $D_{\alpha L}^+$ ($D_{\alpha R}^-$).
$\check m_{D_\alpha}$ is a mass generated by $m_{D_\alpha}$ in (\ref{fermionAction1}).
The matrix $\check \mu$ represents the brane interactions (\ref{braneFmass1}).
Each element $\check \mu_{\alpha\beta}$ is proportional to  
$(\mu^\dagger)_{\alpha\beta} = \mu_{\beta\alpha}^*$.
(Note that $\check \mu$ has dimension of mass and that $\check \mu$ is not proportional to $\mu^\dagger$ as a matrix.)

Mass-eigenstates of up-type quarks are gauge-eigenstates.   However mass-eigenstates of down-type quarks
are not  gauge-eigenstates as a result of $\check \mu$.
${\cal M}_{\rm down}$ can be expressed, in the canonical form, as
\begin{align}
&{\cal M}_{\rm down} = \Omega \,  \begin{pmatrix} M_{\rm down} &\cr & M_D \end{pmatrix} \,  \tilde \Omega^\dagger ~, ~~
\Omega^\dagger = \Omega^{-1} ~,~~ \tilde \Omega^\dagger = \tilde \Omega^{-1} ~,  \cr
\noalign{\kern 5pt}
& M_{\rm down} = \begin{pmatrix} m_d &&\cr & m_s &\cr && m_b \end{pmatrix}, ~~
M_D =  \begin{pmatrix} m_{D_1} &&\cr & m_{D_2} &\cr && m_{D_3} \end{pmatrix}.
\label{downEffMass2}
\end{align}
Note $\Omega \not= \tilde \Omega$ for $\hat \mu \not= 0$.
Mass-eigenstates denoted by $\hat ~$ are given by
\begin{align}
&\begin{pmatrix}  \vec{\hat d}_L  \cr \vec{\hat D}_{L}  \end{pmatrix}
= \Omega^\dagger  \begin{pmatrix}  \vec{d}_L  \cr \vec{D}_L \end{pmatrix} , ~~
\begin{pmatrix}  \vec{\hat d}_R \cr  \vec{\hat D}_R \end{pmatrix}
= \tilde \Omega^\dagger  
\begin{pmatrix}  \vec{d}_R' \cr  \vec{D}_R  \end{pmatrix}, \cr
\noalign{\kern 5pt}
&{\cal L}_m^{\rm down} = - \Big\{ \vec{\bar{\hat d}}_L{}^t M_{\rm down} \,  \vec{\hat d}_R
+  \vec{\bar{\hat D}}_L{}^t M_D \,  \vec{\hat D}_R +  {\rm h.c.} \, \Big\} ~.
\label{downMassEigen1}
\end{align}
All $m_{D_\alpha}$'s are of $O(m_\KK)$, and much larger than $m_d, m_s$ and $m_b$.
Unitary matrices $\Omega$ and $\tilde \Omega$ are decomposed as
\begin{align}
&\Omega = \begin{pmatrix} \Omega_q & \Omega_b \cr \Omega_a & \Omega_D \end{pmatrix} , ~~
\tilde \Omega^\dagger  = 
\begin{pmatrix} \tilde \Omega_q & \tilde \Omega_b \cr \tilde \Omega_a & \tilde \Omega_D \end{pmatrix} , 
\label{unitaryM1}
\end{align}
where all $\Omega_q, \tilde \Omega_q$ etc.\ are 3-by-3 matrices.
The unitarity of $\Omega$  implies that
\begin{alignat}{2}
&\Omega_q \Omega_q^\dagger + \Omega_b \Omega_b^\dagger = I_3 ~, &\quad
&\Omega_q^\dagger \Omega_q +  \Omega_a^\dagger  \Omega_a = I_3 ~, \cr
&\Omega_a \Omega_a^\dagger + \Omega_D \Omega_D^\dagger = I_3 ~, &\quad
&\Omega_b^\dagger \Omega_b +  \Omega_D^\dagger  \Omega_D = I_3 ~, \cr
&\Omega_q \Omega_a^\dagger + \Omega_b \Omega_D^\dagger = 0 ~, &\quad
&\Omega_q^\dagger \Omega_b +  \Omega_a^\dagger  \Omega_D = 0~,
\label{unitarity1}
\end{alignat}
where $I_3$ is a 3-by-3 unit matrix.
Similar relations hold for $\tilde \Omega$.

\subsection{$W$ couplings}

The gauge coupling of $\Psi_{({\bf 3}, {\bf 4})}^\alpha (x,z)$ leads to the $W$ coupling
\begin{align}
{\cal L}_W \simeq \frac{g_L^W}{\sqrt{2}} \, W_\mu  \,  \vec{\bar u}_L \, \Gamma^\mu  \, \vec d_L
+ {\rm h.c.} ~.
\label{Wcoupling1}
\end{align}
In the next section we will confirm that $g_L^W \sim g_w$ and that couplings of right-handed components
are tiny, $g_R^W / g_w \lesssim 10^{-6}$.
It follows from (\ref{downMassEigen1}) that the gauge-eigenstate $\vec{d}_L$ is related 
to the mass-eigenstate $\vec{\hat d}_L $ by 
$\vec{d}_L = \Omega_q \vec{\hat d}_L  + \Omega_b \vec{\hat D}_L$.
For up-type quarks $\vec{u}_L = \vec{\hat u}_L$.
At low energies ($\sqrt{s} \ll m_{D_j}$) the $\hat D$ field may be dropped so that
\begin{align}
{\cal L}_W \simeq \frac{g_L^W}{\sqrt{2}} \, W_\mu  \,  
\vec{\bar{\hat u}}_L \, \Gamma^\mu  \, \Omega_q \vec{\hat d}_L + {\rm h.c.} ~.
\label{Wcoupling2}
\end{align}
In other words the CKM matrix is given by
\begin{align}
V^{\rm CKM} \simeq  \Omega_q ~. 
\label{CKM1}
\end{align}
It should be noted that $\Omega_q$ is not unitary in rigorous sense, as
$\Omega_q \Omega_q^\dagger  = I_3 - \Omega_b \Omega_b^\dagger$.

(\ref{downEffMass1}) and (\ref{downEffMass2}) lead to  
\begin{align}
(q_1):&~
\Omega_q M_{\rm down}  \tilde \Omega_b + \Omega_b  M_D \tilde \Omega_D = 0 ~, \cr
\noalign{\kern 5pt}
(q_2):&~
\Omega_a M_{\rm down}  \tilde \Omega_q + \Omega_D  M_D \tilde \Omega_a = \check \mu ~, \cr
\noalign{\kern 5pt}
(q_3):&~
\Omega_q M_{\rm down}  \tilde \Omega_q + \Omega_b  M_D \tilde \Omega_a = M_{\rm up} ~, \cr
\noalign{\kern 5pt}
(q_4):&~
\Omega_a M_{\rm down}  \tilde \Omega_b + \Omega_D  M_D \tilde \Omega_D = \check m_D ~, 
\label{Omegarelation1}
\end{align}
or equivalently
\begin{align}
(r_1):&~
\Omega_q M_{\rm down} =   M_{\rm up} \, \tilde \Omega_q^\dagger ~, \cr
\noalign{\kern 5pt}
(r_2):&~
\Omega_b M_D =    M_{\rm up}  \,\tilde \Omega_a^\dagger  ~, \cr
\noalign{\kern 5pt}
(r_3):&~
\Omega_a M_{\rm down}  
= \check \mu \, \tilde \Omega_q^\dagger + \check m_D  \, \tilde \Omega_b^\dagger    ~, \cr
\noalign{\kern 5pt}
(r_4):&~
\Omega_D  M_D = \check \mu \,  \tilde \Omega_a^\dagger  + \check m_D \,\tilde \Omega_D^\dagger   ~.
\label{Omegarelation2}
\end{align}
From the relation $(q_1)$ and $(r_2)$ above one finds
\begin{align}
\Omega_b = - \Omega_q M_{\rm down}  \tilde \Omega_b  \tilde \Omega_D^{-1} M_D^{-1} 
= -M_{\rm up} \tilde \Omega_a^\dagger M_D^{-1} ~.
\label{Omegarelation3}
\end{align}
In other words the magnitude of each matrix element of $\Omega_b$, denoted as $|| \Omega_b||$,  is 
\begin{align}
|| \Omega_b || =   O \Big( \frac{m_q}{m_D} \Big) ||  \tilde \Omega_b  || \ll 1 
\label{Omegarelation4}
\end{align}
where $m_q = m_d, m_s, m_b$ and $m_D = m_{D_j}$.  As $m_b/m_D \sim 10^{-3}$, 
$\Omega_q$ is nearly unitary.
As $\Omega_a = - (\Omega_D^\dagger)^{-1} \Omega_b^\dagger \Omega_q$, one sees
that $|| \Omega_a || = O(m_q/m_D)$ as well.

\ignore{
The relation $(q_3)$ becomes, with the aid of (\ref{unitarity1}) and (\ref{Omegarelation3}), 
\begin{align}
\Omega_q M_{\rm down} \tilde \Omega_q ~ \frac{1}{I_3 - \tilde \Omega_a^\dagger \tilde \Omega_a}
= \Omega_q M_{\rm down} (\tilde \Omega_q^\dagger  )^{-1} = M_{\rm up} ~, 
\label{Massrelation2}
\end{align}
or
\begin{align}
\Omega_q M_{\rm down} = M_{\rm up} \tilde \Omega_q^\dagger ~.
\label{Massrelation3}
\end{align}
}

Further $(q_2)$ and $(q_4)$ in (\ref{Omegarelation1})  imply that
\begin{align}
\check \mu \sim \Omega_D M_D  \tilde \Omega_a ~, ~~ 
\check m_D \sim \Omega_D M_D \tilde \Omega_D ~.
\label{Massrelation1}
\end{align}
The relation $(r_1)$ in (\ref{Omegarelation2}) gives a severe constraint on the mass spectrum.  
Recall (\ref{CKM1}), which implies that
\begin{align}
\frac{m_{dk}}{m_{uj}} \, |V^{\rm CKM}_{jk}| \sim | ( \tilde \Omega_q^\dagger )_{jk} |  < 1  
\label{CKM2}
\end{align}
where $(m_{d1}, m_{d2}, m_{d3}) = (m_d, m_s, m_b)$ and $(m_{u1}, m_{u2}, m_{u3}) = (m_u, m_c, m_t)$.
The observed mean value (magnitude) of $V^{\rm CKM}$ is
\begin{align}
V^{\rm CKM}_{\rm obs} \sim 
\begin{pmatrix} 0.974 & 0.224 & 0.004 \cr 0.218 & 0.997 & 0.042 \cr 0.008 & 0.030 & 1.019 \end{pmatrix} .
\label{CKM3}
\end{align}
The observed $m_u \sim 1.3\,$MeV is too small, and the inequality (\ref{CKM2}) is not satisfied for
the 11, 12 and 13 elements.  Rigorous treatment presented in the previous and next sections also 
confirms this behavior.
In the present paper we tentatively suppose that $m_u \sim 20\,$MeV.  
The issue of small $m_u$ is left for future investigation.

\subsection{$Z$ couplings}
For up-type quarks one finds 
\begin{align}
{\cal L}_Z^{\rm up} \sim  - \frac{g_w}{\cos \theta_W} \, Z_\mu 
\bigg\{ \frac{1}{2} \, \vec{\bar{\hat u}}_L \Gamma^\mu \vec{\hat u}_L
- \frac{2}{3} \sin^2 \theta_W \Big( \vec{\bar{\hat u}}_L \Gamma^\mu \vec{\hat u}_L
+ \vec{\bar{\hat u}}_R \Gamma^\mu \vec{\hat u}_R \Big) \bigg\} ~.
\label{Zcoupling1}
\end{align}
Recall that $D_\alpha$ fields are $SO(5)$ singlet.
$Z$ couplings of  down-type quarks are given by
\begin{align}
{\cal L}_Z^{\rm down} \sim  &- \frac{g_w}{\cos \theta_W} \, Z_\mu 
\bigg\{ - \frac{1}{2} \, \vec{\bar{d}}_L \Gamma^\mu \vec{d}_L \cr
\noalign{\kern 5pt}
&+ \frac{1}{3} \sin^2 \theta_W \Big( \vec{\bar{d}}_L \Gamma^\mu \vec{d}_L
+  \vec{\bar{D}}_L \Gamma^\mu \vec{D}_L+ \vec{\bar{d'}}_R \Gamma^\mu \vec{d}_R'  
+ \vec{\bar{D}}_R \Gamma^\mu \vec{D}_R\Big) 
\bigg\} ~.
\label{Zcoupling2}
\end{align}
In terms of mass-eigenstates in (\ref{downMassEigen1}), $Z$ couplings at low energies are expressed as
\begin{align}
{\cal L}_Z^{\rm down} 
&\sim  - \frac{g_w}{\cos \theta_W} \, Z_\mu   \bigg\{-\frac{1}{2} \, 
\Big( \vec{\bar{\hat d}}_L{}^t \Omega_q^\dagger + \vec{\bar{\hat D}}_L{}^t \Omega_b^\dagger \Big)
 \Gamma^\mu 
 \Big( \Omega_q \vec{\hat d}_L + \Omega_b \vec{\hat D}_L \Big)  \cr
\noalign{\kern 5pt}
&\quad 
+ \frac{1}{3} \sin^2 \theta_W \Big( \vec{\bar{\hat d}}_L \Gamma^\mu \vec{\hat d}_L
+  \vec{\bar{\hat D}}_L \Gamma^\mu \vec{\hat D}_L
+ \vec{\bar{\hat d}}_R \Gamma^\mu \vec{\hat d}_R  
+ \vec{\bar{\hat D}}_R \Gamma^\mu \vec{\hat D}_R\Big)    \bigg\}  \cr
\noalign{\kern 5pt}
&\sim  - \frac{g_w}{\cos \theta_W} \, Z_\mu   \bigg\{ -\frac{1}{2} \, 
\vec{\bar{\hat d}}_L\Gamma^\mu  \Omega_q^\dagger \Omega_q \vec{\hat d}_L
+ \frac{1}{3} \sin^2 \theta_W \Big( \vec{\bar{\hat d}}_L \Gamma^\mu \vec{\hat d}_L
+ \vec{\bar{\hat d}}_R \Gamma^\mu \vec{\hat d}_R   \Big) \bigg\} .
\label{Zcoupling3}
\end{align}
In the first term $ \Omega_q^\dagger \Omega_q = I_3 - \Omega_a^\dagger \Omega_a$, 
and the $\Omega_a^\dagger \Omega_a$ term gives rise to  FCNC.
However, with the use of the last two relations in (\ref{unitarity1}) and 
the relation (\ref{Omegarelation3}) one sees
\begin{align}
\Omega_a^\dagger \Omega_a 
= \Omega_q^\dagger \Omega_b \Omega_b^\dagger (\Omega_q^\dagger)^{-1}
= O \Big( \frac{m_q^2}{m_D^2} \Big) \lesssim 10^{-6} ~.
\label{FCNC1}
\end{align}
FCNC interactions are  naturally suppressed.
The FCNC suppression will be confirmed by rigorous treatment in the next section as well.

\section{Evaluation of gauge couplings}

In section 3 we have obtained wave functions of gauge bosons and quarks, with which 
gauge couplings of quarks can be evaluated.  
Given the parameters $\mu_{\alpha \beta}$ of the brane interaction (\ref{braneFmass1})
and the Dirac masses $m_{D_\alpha}$ for the $D_\alpha^\pm$ fields, 
the bulk mass parameters $c_{D_\alpha}$ are chosen such that the mass spectrum 
of down-type quarks are reproduced by the condition (\ref{downmass2}).
Then the wave functions of all quarks are unambiguously determined.
The parameters $\mu_{\alpha \beta}$ need be chosen such that the observed
CKM mixing matrix is reproduced.  This process, however,  is not so trivial.

As inferred in the effective theory formulated in the previous section, 
consistent solutions are available only when $m_d < m_u$.  
This behavior has been already recognized in the case of no-mixing in ref.\ \cite{FHHOY2019a}.
In this section we present the detailed results for the $W$ and $Z$ couplings of quarks
with typical $\mu_{\alpha \beta}$.  It will be seen that a simple form of $\mu$ matrix leads to
reasonable  CKM mixing matrix, though it may not be perfect.

$m_Z$, $z_L = 10^{10}$, $m_t$, $m_b$, $m_c$, $m_s$,  $m_u$, and $m_d$ are inputs.
The bare Weinberg angle $\sin^2 \theta_W^0 = s_\phi^2/(1+ s_\phi^2)$ 
with a given $\theta_H$  is determined  to fit the LEP1 data for $e^+ e^- \go \mu^+ \mu^-$ 
at $\sqrt{s} = m_Z$.\cite{LEPsummary}
It will be seen below that evaluated gauge couplings turn out very close to those in the SM
with $\sin^2 \theta_W = 0.2312$.
The values for $m_\KK$, $c_u$, $c_c$, $c_t$ etc.\ with given $\theta_H$ are 
summarized in Table \ref{Tab:cqMkk}.

\begin{table}[tbh]
\renewcommand{\arraystretch}{1.2}
\begin{center}
\caption{Values of $m_\KK$, $k$, $\sin^2 \theta_W^0 = s_\phi^2/(1+ s_\phi^2)$, 
$c_u$, $c_c$, $c_t$ are tabulated for $\theta_H = 0.10, ~0.15$ and $z_L = 10^{10}$.
We set $m_Z = 91.1876\,$GeV, $\alpha_\EM (m_Z) = 1/128$  and 
$(m_u, m_c, m_t) = (0.020, 0.619, 171.17)\,$GeV.
The value $m_u > m_d$ has been used for a reason explained in the text.
}
\vskip 10pt
\begin{tabular}{|c|c|c|c|c|c|c|}
\hline
$\theta_H$ & $\mymat{m_\KK}{{\rm (TeV)}}$ & $\mymat{k}{{\rm (GeV)}}$ & $\sin^2 \theta_W^0$ &$c_u$ &$c_c$ &$c_t$\\
\hline
$0.10$ &$12.08$ &$3.84 \times 10^{13}$ &$0.2306$ &$- 0.9169$ &$- 0.7545$ &$- 0.2274$\\
\hline
$0.15$ &$8.07$ &$2.57 \times 10^{13}$ &$0.2299$ &$- 0.9170$ &$- 0.7546$ &$- 0.2294$\\
\hline 
\end{tabular}
\label{Tab:cqMkk}
\end{center}
\end{table}

In general nine elements of the brane interaction matrix $\mu$ can be complex.
Six out of nine  phases  can be absorbed by redefinition of the  fields $\vec d_R'$ and $\vec D_L^+$.
Three of them remain as CP violation phases.   When all heavy fields such as $\vec D^\pm$ are
integrated out, only one complex phase survives at the CKM matrix level.
In the present paper we consider a real matrix $\mu$, which is parametrized as
\begin{align}
\mu = U_{12} (\phi_{12}) U_{13} (\phi_{13}) U_{23} (\phi_{23}) 
\begin{pmatrix} \mu_1 &&\cr & \mu_2 &\cr &&\mu_3 \end{pmatrix}
U_{23} (\omega_{23})^\dagger U_{13} (\omega_{13})^\dagger U_{12} (\omega_{12})^\dagger ~.
\label{mumatrix1}
\end{align}
Here $U_{jk} (\phi)$ is a rotation matrix in the $jk$ subspace;
\begin{align}
U_{12} (\phi) = 
\begin{pmatrix} \cos \phi & \sin \phi & 0 \cr - \sin\phi & \cos \phi & 0 \cr 0 & 0 & 1 \end{pmatrix} .
\end{align}
As typical values we set $\tilde m_{D_d} = \tilde m_{D_s} = \tilde m_{D_b} = 1$.
For the $\mu$ matrix, we take  $(\mu_1, \mu_2, \mu_3) = (0.1, 0.1, 1)$ as reference
values suggested in ref.\ \cite{FHHOY2019a}.
Among the rotation angles in (\ref{mumatrix1}), $\omega_{12}$ is most responsible
for the Cabibbo angle.  We have explored the parameter space $(\omega_{12}, \omega_{23})$,
while keeping $\phi_{jk} = \omega_{13} = 0$.
Given $\mu$, the bulk mass parameters of $D$ fields, $(c_{D_d}, c_{D_s}, c_{D_b})$, are
determined so as to reproduce $(m_d, m_s, m_b)$.
This turns out possible only for appropriate $\mu$.
With all the parameters set, wave functions of down-type quarks are determined, and
gauge couplings of quarks are evaluated.
Sets of typical values of these parameters are tabulated in Table \ref{Tab:cDmu}. 
We note that the masses of the first KK excited states of $d, s, b$ quarks turn out
around $0.6 \, m_\KK$.

\begin{table}[tbh]
\renewcommand{\arraystretch}{1.2}
\begin{center}
\caption{Sets of parameters which yield a reasonable CKM matrix.
$(c_{D_d}, c_{D_s}, c_{D_b})$ is determined to give
$(m_d, m_s, m_b) = (0.0029, 0.055, 2.89)\,$GeV by (\ref{downmass2}).
We set $\phi_{jk} = \omega_{13} = 0$ in (\ref{mumatrix1}) and 
$\tilde m_{D_d} = \tilde m_{D_s} = \tilde m_{D_b} = 1$.
}
\vskip 10pt
\begin{tabular}{|c|c|c|c|c|c|c|}
\hline
&$\theta_H$ & $(\mu_1, \mu_2, \mu_3)$ & $(\omega_{12}, \omega_{23})$ 
&$c_{D_d}$ &$c_{D_s}$ &$c_{D_b}$\\
\hline
(a) & $0.10$ &$(0.1, 0.1, 1)$ &$(0.1055, 0.0018)$ &$0.520074$ &$0.751360$ &$0.951239$ \\
\hline
(b) &$0.15$ &$(0.1, 0.1, 1)$ &$(0.1055, 0.00198)$  &$0.478059$ &$ 0.751545$ &$0.955367$\\
\hline 
\end{tabular}
\label{Tab:cDmu}
\end{center}
\end{table}

Wave functions of each down-type quark consist of 12 components, $(d, d', D_d^+, D_d^-)$, 
 $(s, s', D_s^+, D_s^-)$,  $(b, b', D_b^+, D_b^-)$.
Coefficient vectors,   $\vec \alpha$, $\vec \alpha '$, $\vec a$ and $\vec b$
in (\ref{down-wave1})  and (\ref{coefficient1})   for $\theta_H = 0.15$
with the parameter set (b) in Table \ref{Tab:cDmu} are tabulated in Table \ref{Tab:waveF1}.
With these coefficients wave functions of left- and right-handed components, $f_{jL} (z)$ and
$f_{jR} (z)$, are determined.  
In Table \ref{Tab:norm1} norm of each component $N_{jL/R} =  \int_1^{z_L} dz \, |f_{jL/R}|^2$ 
is listed.   Note $\sum_j N_{jL} = \sum_j N_{jR} = 1$.

\begin{table}[tbh]
\renewcommand{\arraystretch}{1.2}
\begin{center}
\caption{Coefficient vectors  in (\ref{coefficient1}) for
wave functions of down-type quarks for $\theta_H = 0.15$
with the parameter set (b) in Table \ref{Tab:cDmu} are listed.
$(\hat d, \hat s, \hat b)$ represent mass-eigenstates. 
}
\vskip 10pt
\begin{tabular}{|c||c|c|c|c|}
\hline
&$\vec \alpha$ &$\vec \alpha '$ &$\vec a$ &$\vec b$ \\
\hline
&$1.640$ &$-8.692 \times 10^{-6} \, i$ &~$0.007734 \, i$~  &$2.207 \times 10^{-9} \, i$\\
$\hat d$ & $-0.3588$ &$2.148 \times 10^{-6} \, i$ &$-0.4697 \, i$ &$-1.178 \times 10^{-7} \, i$\\
&$1.476 \times 10^{-5}$ &$- 1.520 \times 10^{-10} \, i$ &$0.005361 \, i$  &$1.232 \times 10^{-9} \, i$ \\
\hline 
&$0.3812$  &$- 3.832 \times 10^{-5} \, i$ &$0.03452 \, i$ &$1.868 \times 10^{-7} \, i$\\
$\hat s$ &$1.542$ &$- 1.752 \times 10^{-4} \, i$ &$0.04968\, i $ &$2.362 \times 10^{-7} \, i$ \\
&$-0.02291$ &$4.474 \times 10^{-6} \, i $ &$-0.4376 \, i$ &$-1.908 \times 10^{-6} \, i$\\
\hline 
 &$0.007211$ &$- 3.809 \times 10^{-5}\, i $ &$0.03431 \, i$ &$9.756 \times 10^{-6} \, i$\\
$\hat b$ &$0.02927$ &$- 1.746 \times 10^{-4}\, i $ &$0.05525 \, i$ &$1.380 \times 10^{-5} \, i$\\
&$1.208$ &$- 0.01239 \, i $ &$0.4389 \, i$   &$1.005 \times 10^{-4} \, i$\\
\hline
\end{tabular}
\label{Tab:waveF1}
\end{center}
\end{table}
\begin{table}[tbh]
\renewcommand{\arraystretch}{1.2}
\begin{center}
\caption{Norm of each component of down-type quarks for $\theta_H = 0.15$
with the parameter set (b) in Table \ref{Tab:cDmu} is listed.
$(\hat d, \hat s, \hat b)$ represent mass-eigenstates. 
In this table $10^{-13}$, for instance, implies order of $10^{-13}$.
}
\vskip 10pt
\begin{tabular}{|c||c|c|c||c|c|c|}
\hline
&$\hat d_L$ &$\hat s_L$ &$\hat b_L$ &$\hat d_R$ &$\hat s_R$ &$\hat b_R$\\
\hline
$d$ &$0.9487$ &$0.0513$ &$10^{-5}$  &$0.0001$  &$0.0022$&$0.0022$\\
$s$ &$0.0513$ &$0.9484$ &$0.0003$ &$10^{-8}$  &$10^{-5}$ &$10^{-5}$\\
$b$ &$10^{-10}$ &$0.0004$ &$0.9996$  &$10^{-22}$  &$10^{-13}$  &$10^{-6}$\\
\hline
$d'$ &$10^{-23}$  &$10^{-19}$ &$10^{-15}$ &$0.0198$&$0.3856$ &$0.3810$\\
$s'$ &$10^{-24}$ &$10^{-18}$ &$10^{-14}$ &$10^{-6}$ &$0.0074$ &$0.0074$\\
$b'$ &$10^{-32}$ &$10^{-21}$ &$10^{-10}$ &$10^{-20}$ &$10^{-11}$ &$0.0003$\\
\hline
$D_d^+$ &$10^{-17}$ &$10^{-13}$ &$10^{-9}$ &$0.0023$ &$0.0465$ &$0.0459$\\
$D_s^+$ &$10^{-13}$ &$10^{-12}$ &$10^{-9}$ &$0.3113$ &$0.0035$ &$0.0043$\\
$D_b^+$ &$10^{-17}$ &$10^{-10}$ &$10^{-7}$   &$10^{-5}$&$0.1177$&$0.1184$\\
\hline
$D_d^-$ &$10^{-17}$ &$10^{-13}$ &$10^{-9}$    &$0.0025$  &$0.0489$ &$0.0484$\\
$D_s^-$ &$10^{-13}$ &$10^{-13}$ &$10^{-9}$ &$0.6639$ &$0.0074$ &$0.0092$\\
$D_b^-$ &$10^{-17}$ &$10^{-11}$ &$10^{-7}$ &$0.0001$ &$0.3808$ &$0.3830$\\
\hline
\end{tabular}
\label{Tab:norm1}
\end{center}
\end{table}

It is seen that the left-handed components of mass-eigenstates,  $(\hat d_L, \hat s_L, \hat b_L)$,
are mostly contained in the original $(d, s, b)$ fields.   
On the other hand the right-handed components $(\hat d_R, \hat s_R, \hat b_R)$
are distributed among various components.  
Dominant parts of $\hat d_R$ are in $D_s^\pm$, $\hat s_R$ in $d'$, $D_d^\pm$ and $D_b^\pm$, 
and $\hat b_R$ in $d'$, $D_d^\pm$ and $D_b^\pm$.
The pattern of distribution for the right-handed components depends on the form of 
the brane interaction, or on the $\mu$ matrix.
A crucial  point is that $d'$ component of $\hat d_R$, $s'$ component of $\hat s_R$, 
and $b'$ component of $\hat b_R$ are all small.  As is seen in the following  subsection, 
this property is important to assure vanishingly small $W$ couplings of right-handed quarks.

\subsection{$W$ couplings}

The $SO(5)$ gauge potentials can be expanded as
\begin{align}
A_M  = \sum_{a=1}^3 \Big\{ A_M^{a_L} T^{a_L} + A_M^{a_R} T^{a_R} + A_M^{\hat a} T^{\hat a} \Big\}
+ A_M^{\hat 4} T^{\hat 4} ~,
\label{SO5pot1}
\end{align}
where $T^{a_L}$ and $T^{a_R}$ are $SU(2)_L$ and $SU(2)_R$ generators, respectively.
$\{ T^{\hat p} ; p=1, \cdots, 4\}$ are generators of $SO(5)/SO(4)$.
In the spinor representation, for instance, 
\begin{align}
&T^{a_L} = \frac{1}{2} \begin{pmatrix} \sigma^a & 0 \cr 0 & 0 \end{pmatrix} ,  ~~
T^{a_R} = \frac{1}{2} \begin{pmatrix} 0 & 0 \cr 0 &  \sigma^a  \end{pmatrix} ,  \cr
\noalign{\kern 5pt}
&T^{\hat a} =  \frac{1}{2\sqrt{2}}  \begin{pmatrix} 0 & i \sigma^a \cr -i \sigma^a & 0 \end{pmatrix}, ~
T^{\hat 4} = \frac{1}{2\sqrt{2}}  \begin{pmatrix} 0 & I_2 \cr I_2 & 0 \end{pmatrix}
\label{so5generator1}
\end{align}
where $\sigma^a$'s and $I_2$ are Pauli matrices and a 2-by-2 unit matrix.
$W$ boson is contained,  in the twisted gauge, in 
\begin{align}
\tilde A_\mu &\Rightarrow \frac{1}{2} \Big\{
(\tilde A_\mu^{1_L} - i \tilde A_\mu^{2_L}) (T^{1_L} + i T^{2_L})
+ (\tilde A_\mu^{1_R} - i \tilde A_\mu^{2_R}) (T^{1_R} + i T^{2_R}) \cr
\noalign{\kern 5pt}
&\hskip 2.cm
+  (\tilde A_\mu^{\hat 1} - i \tilde A_\mu^{\hat 2}) (T^{\hat 1} + i T^{\hat 2}) \Big\} + ~ {\rm h.c.} \cr
\noalign{\kern 5pt}
&\Rightarrow \frac{1}{2} \Big\{
(1 + c_H)  \mathring{W}_\mu (T^{1_L} + i T^{2_L}) 
+ (1 - c_H)  \mathring{W}_\mu (T^{1_R} + i T^{2_R}) \cr
\noalign{\kern 5pt}
&\hskip 2.cm 
- \sqrt{2} \, s_H \mathring{W}_\mu^S (T^{\hat 1} + i T^{\hat 2}) \Big\} + ~ {\rm h.c.} ~.
\label{Wint1}
\end{align}
Here the expression  (\ref{waveGauge1}) has been inserted.
$W$ couplings of quarks come solely from the couplings of $\Psi_{({\bf 3}, {\bf 4})}^\alpha$.
\begin{align}
{\cal L}_W^{d=4} &= - ig_A \int_1^{z_L} \frac{dz}{k} \bigg\{
\mathring{W}_\mu  \bigg( \frac{1+c_H}{2} \,  \vec{\bar{\tilde u}} \,  \Gamma^\mu \, \vec{\tilde d}
+ \frac{1- c_H}{2} \,  \vec{\bar{\tilde u}}' \,  \Gamma^\mu \, \vec{\tilde d}' ~ \bigg) \cr
\noalign{\kern 5pt}
&\hskip 2.cm
+ \mathring{W}_\mu^S \bigg( - i \frac{s_H}{2} \, \vec{\bar{\tilde u}} \,  \Gamma^\mu \, \vec{\tilde d}'
+ i \frac{s_H}{2} \, \vec{\bar{\tilde u}}' \,  \Gamma^\mu \, \vec{\tilde d} ~\bigg) 
 \bigg\}  + ~ {\rm h.c.} ~ .
\label{Wcoupling4D1}
\end{align}
Here, as in (\ref{notation1}), we have denoted as
\begin{align}
\vec u = \begin{pmatrix} \check u \cr \check c \cr \check t \end{pmatrix}, ~
\vec u' = \begin{pmatrix} \check u' \cr \check c' \cr \check t' \end{pmatrix} .
\label{notation2}
\end{align}

We use the following notation for wave functions of quarks.  4D quark fields are
denoted by hat $\hat{~}$;
\begin{align}
\begin{pmatrix} \hat u_1(x)  \cr \hat u_2(x)  \cr  \hat u_3(x)  \end{pmatrix}  = 
\begin{pmatrix} \hat u(x)  \cr \hat c(x)  \cr  \hat t(x)  \end{pmatrix} , ~
\begin{pmatrix} \hat d_1(x)  \cr \hat d_2(x)  \cr  \hat d_3(x)  \end{pmatrix}  = 
\begin{pmatrix} \hat d(x)  \cr \hat s(x)  \cr  \hat b(x)  \end{pmatrix} .
\label{notation3}
\end{align}
For up-type quarks 5D fields in the twisted gauge are expanded as
\begin{align}
\tilde{\check u}_j (x,z) &= \sqrt{k} \, \Big\{ \hat u_{jL} (x) f^{\hat u_j}_{L u_j} (z) 
+  \hat u_{jR} (x) f^{\hat u_j}_{R u_j} (z)  \Big\} ~, \cr
\noalign{\kern 5pt}
\tilde{\check u}_j' (x,z) &= \sqrt{k} \, \Big\{ \hat u_{jL} (x) f^{\hat u_j}_{L u_j'} (z) 
+  \hat u_{jR} (x) f^{\hat u_j}_{R u_j'} (z)  \Big\} ~.
\label{notationUp1}
\end{align}
With the expression in (\ref{waveUp1}), for instance, 
\begin{align}
f^{\hat u_1}_{L u_1} (z)  &= \bar c_H C_L(z; \lambda_u, c_u) / \sqrt{r_u} ~, \cr
\noalign{\kern 5pt}
f^{\hat u_2}_{L u_2'} (z)  &=  i  \bar s_H \hat S_L(z; \lambda_c, c_c)/ \sqrt{r_c} ~.
\label{notationUp2}
\end{align}
For down-type quarks 5D fields in the twisted gauge are expanded as
\begin{align}
\tilde{\check d}_j (x,z) &= \sqrt{k} \, \sum_{m=1}^3 \Big\{ \hat d_{mL} (x) f^{\hat d_m}_{L d_j} (z) 
+  \hat d_{mR} (x) f^{\hat d_m}_{R d_j} (z)  \Big\} ~, \cr
\noalign{\kern 3pt}
\tilde{\check d}_j' (x,z) &= \sqrt{k} \, \sum_{m=1}^3 \Big\{ \hat d_{mL} (x) f^{\hat d_m}_{L d_j'} (z) 
+  \hat d_{mR} (x) f^{\hat d_m}_{R d_j'} (z)  \Big\} ~,  \cr
\noalign{\kern 3pt}
{\check D}_j^+ (x,z) &= \sqrt{k} \, \sum_{m=1}^3 \Big\{ \hat d_{mL} (x) f^{\hat d_m}_{L D_j^+} (z) 
+  \hat d_{mR} (x) f^{\hat d_m}_{R D_j^+} (z)  \Big\} ~, \cr
\noalign{\kern 3pt}
{\check D}_j^- (x,z) &= \sqrt{k} \, \sum_{m=1}^3 \Big\{ \hat d_{mL} (x) f^{\hat d_m}_{L D_j^-} (z) 
+  \hat d_{mR} (x) f^{\hat d_m}_{R D_j^-} (z)  \Big\} ~.
\label{notationDown1}
\end{align}
With the expression in (\ref{down-wave1}), one finds, for instance,
\begin{align}
f^{\hat d_1}_{L d_2} (z) &= \alpha_s^{\hat d} C_L(z; \lambda_d, c_c) ~, \cr
\noalign{\kern 5pt}
f^{\hat d_2}_{R d_3'} (z) &= \alpha_{b'}^{\hat s} C_R(z; \lambda_s, c_t) ~, \cr
\noalign{\kern 5pt}
f^{\hat d_3}_{R D_1^+} (z) &=
a_{d}^{\hat b} {\cal S}_{R2}(z;\lambda_b, c_{D_d}, \tilde m_{D_d})
+b_{d}^{\hat b}  {\cal S}_{R1}(z;\lambda_b, c_{D_d}, \tilde m_{D_d} ) ~.
\label{notationDown2}
\end{align}
Here  $\vec \alpha^{\hat d_j}$, $\vec \alpha'{}^{\hat d_j}$, 
$\vec a^{\hat d_j}$ and $\vec b^{\hat d_j}$ are the coefficient vectors
determined for $\hat d_j$.

$W$ couplings of quarks are defined by
\begin{align}
{\cal L}_W^{d=4} &=\frac{i}{\sqrt{2}} W_\mu \sum_{j,k} \Big\{ 
g_{Ljk}^W \, \bar{\hat u}_{jL} \Gamma^\mu \hat d_{kL} + 
g_{Rjk}^W \, \bar{\hat u}_{jR} \Gamma^\mu \hat d_{kR} \Big\} + ~ {\rm h.c.} ~.
\label{Wcoupling4D2}
\end{align}
Inserting (\ref{notationUp1}) and (\ref{notationDown1}) into (\ref{Wcoupling4D1}), 
one finds
\begin{align}
&\begin{bmatrix} g_{Ljk}^W \cr \noalign{\kern 3pt} g_{Rjk}^W \end{bmatrix} 
= - ig_w \frac{\sqrt{kL}}{\sqrt{r_W}} \int_1^{z_L} dz   \cr
\noalign{\kern 5pt}
&
\times \Bigg\{ C(z, \lambda_W) \Bigg(
\frac{1+c_H}{2} \begin{bmatrix}  f^{\hat u_j}_{L u_j} (z)^*  f^{\hat d_k}_{L d_j} (z) \cr
f^{\hat u_j}_{R u_j} (z)^*  f^{\hat d_k}_{R d_j} (z) \end{bmatrix} 
+ \frac{1-c_H}{2} \Bigg[ \begin{matrix}  f^{\hat u_j}_{L u_j'} (z)^*  f^{\hat d_k}_{L d_j'} (z) \cr
f^{\hat u_j}_{R u_j'} (z)^*  f^{\hat d_k}_{R d_j'} (z) \end{matrix} \Bigg] \Bigg) \cr
\noalign{\kern 5pt}
&\hskip 1.0cm 
+ \hat S(z, \lambda_W) (-i) \frac{s_H}{2} \Bigg[ 
\begin{matrix}  f^{\hat u_j}_{L u_j} (z)^*   f^{\hat d_k}_{L d_j'} (z) 
- f^{\hat u_j}_{L u_j'} (z)^*   f^{\hat d_k}_{L d_j} (z) \cr
f^{\hat u_j}_{R u_j} (z)^*   f^{\hat d_k}_{R d_j'} (z) 
- f^{\hat u_j}_{R u_j'} (z)^*   f^{\hat d_k}_{R d_j} (z) \end{matrix} \Bigg] \Bigg\} ~.
\label{Wcoupling4D3}
\end{align}
Let us denote the couplings in the matrix form; 
$(\widehat g_L^W)_{jk} = g_{Ljk}^W$ and $(\widehat g_R^W)_{jk} = g_{Rjk}^W$.  
$\widehat  g_L^W$ is parametrized as
\begin{align}
\widehat g_L^W = g_L^W \, \widehat  V_{\rm CKM} ~, ~~ \det V_{\rm CKM} = 1 ~.
\label{Wcoupling4D4}
\end{align}
$\widehat  g_L^W$ and $\widehat  g_R^W$ are evaluated  for the two sets of parameters
in Table \ref{Tab:cDmu};
\begin{align}
\hbox{(a) ~} &\theta_H = 0.10 ~: \cr
\noalign{\kern 5pt}
&g_L^W = 0.9978 \, g_w ~, ~~ 
\widehat  V_{\rm CKM} = \begin{pmatrix} 0.9744 & 0.2245 & 0.0031 \cr
-0.2245 &0.9743 &0.0134 \cr 9 \times 10^{-6} &	- 0.0138 &1.0002 \end{pmatrix} , \cr
\noalign{\kern 5pt}
&\widehat g_R^W = g_w \begin{pmatrix} 2 \times 10^{-12} & 8 \times 10^{-12} & 6 \times 10^{-12} \cr
-1 \times 10^{-11} & 9 \times  10^{-10} & 7 \times 10^{-10} \cr
1 \times 10^{-13} &- 3 \times 10^{-9} & 1 \times 10^{-5} \end{pmatrix} , \cr
\noalign{\kern 10pt}
\hbox{(b) ~} &\theta_H = 0.15 ~: \cr
\noalign{\kern 5pt}
&g_L^W = 0.9950 \, g_w ~, ~~ 
\widehat  V_{\rm CKM} = \begin{pmatrix} 0.9737 & 0.2264 & 0.0043 \cr
-0.2264 &0.9736 &0.0185 \cr 1 \times 10^{-5} &	- 0.0190 &1.0004 \end{pmatrix} , \cr
\noalign{\kern 5pt}
&\widehat g_R^W = g_w \begin{pmatrix} 4 \times 10^{-12} & 1 \times 10^{-11} & 2 \times 10^{-11} \cr
-3 \times 10^{-11} & 2 \times  10^{-9} & 2 \times 10^{-9} \cr
4 \times 10^{-13} &- 1 \times 10^{-8} & 3 \times 10^{-5} \end{pmatrix} .
\label{Wcoupling4D5}
\end{align}
We have checked remarkable cancellation among four terms in the right-handed couplings 
$g_{Rjk}^W$ in (\ref{Wcoupling4D3}).
The resultant $\widehat  V_{\rm CKM}$ is reasonably close to the observed CKM matrix, 
although the 31 element is still too small.
We have evaluated the $W$ couplings of leptons as well.  The couplings of left-handed leptons
$(e, \mu, \tau)$ are  $(0.997665, 0.997662, 0.997659) g_w$ for $\theta_H =0.10$, 
and $(0.994756, 0.994748, 0.994743) g_w$ for $\theta_H =0.15$.  
The relative coupling $g_L^W$ to $g_{L\,  \rm lepton}^W$ is
$g_L^W/g_{L\,  \rm lepton}^W  = 1.00013$ and $1.00028$ for $\theta_H = 0.10$ and $0.15$,
respectively.  The universality holds to high accuracy.  
The $W$ couplings of right-handed  leptons are typically of order $10^{-20} \, g_w$.

\subsection{$Z$ couplings}

Photon $\gamma$ and $Z$ boson are contained in
\begin{align}
&\tilde A_\mu + \frac{g_B}{g_A} \, Q_X B_\mu \Rightarrow 
 ( \tilde A_\mu^{3_L} T^{3_L} + \tilde A_\mu^{3_R} T^{3_R} 
+\tilde A_\mu^{\hat 3} T^{\hat 3} ) + \frac{g_B}{g_A}\,  Q_X B_\mu \cr
\noalign{\kern 5pt}
&\Rightarrow  \frac{\sqrt{\smash[b]{1+s_\phi^2}}}{\sqrt{2}} \, \Big\{
\big[ (1+c_H) T^{3_L} +  (1-c_H) T^{3_R} \big] \mathring Z_\mu
- \sqrt{2} s_H T^{\hat 3} \mathring Z_\mu^S \Big\} \cr
\noalign{\kern 5pt}
&\hskip 2.cm
+  \frac{s_\phi}{\sqrt{\smash[b]{1+s_\phi^2}}} \, Q_\EM
(\mathring A_\mu^\gamma - \sqrt{2} s_\phi \mathring Z_\mu ) ~.
\label{Zcoupling4D1}
\end{align}
Here (\ref{waveGauge1}) and the relation $Q_\EM = T^{3_L} +  T^{3_R} + Q_X$
have been used.
Photon couplings are given by
\begin{align}
{\cal L}_\gamma^{d=4} &= - ig_A \frac{s_\phi}{\sqrt{\smash[b]{1+s_\phi^2}}}  
\int_1^{z_L} \frac{dz}{k} \, \mathring A_\mu^\gamma \bigg\{
\frac{2}{3} \Big( \vec{\bar{\tilde u}} \,  \Gamma^\mu \, \vec{\tilde u} + 
\vec{\bar{\tilde u}}' \,  \Gamma^\mu \, \vec{\tilde u}' \Big) \cr
\noalign{\kern 5pt}
&- \frac{1}{3} \Big( \vec{\bar{\tilde d}} \,  \Gamma^\mu \, \vec{\tilde d} + 
\vec{\bar{\tilde d}}' \,  \Gamma^\mu \, \vec{\tilde d}' +
\vec{\bar{D}}^+ \,  \Gamma^\mu \, \vec{D}^+ 
+ \vec{\bar{D}}^- \,  \Gamma^\mu \, \vec{D}^- \Big) \bigg\} ~.
\label{photoncoupling4D1}
\end{align}
Inserting (\ref{tower1}),   (\ref{notationUp1}) and (\ref{notationDown1}) into (\ref{photoncoupling4D1}),
one finds that
\begin{align}
{\cal L}_\gamma^{d=4} &= - i g_w \frac{s_\phi}{\sqrt{\smash[b]{1+s_\phi^2}}}  
A_\mu^\gamma (x) \int_1^{z_L} dz \, J_\gamma^\mu (x,z) ~, \cr
\noalign{\kern 5pt}
J_\gamma^\mu (x,z) &= \frac{2}{3} \sum_{j =1}^3\Big[ 
\bar{\hat u}_{jL}\Gamma^\mu \hat u_{jL} (x) \Big\{ f^{\hat u_j}_{L u_j} (z)^*  f^{\hat u_j}_{L u_j} (z) 
+ f^{\hat u_j}_{L u_j'} (z)^*  f^{\hat u_j}_{L u_j'} (z)  \Big\} + (L \go R) \Big] \cr
\noalign{\kern 5pt}
& - \frac{1}{3} \sum_{\ell,m=1}^3 \Big[ \bar{\hat d}_{\ell L}\Gamma^\mu \hat d_{mL} (x)
\sum_{j=1}^3 \Big\{ f^{\hat d_\ell}_{L d_j} (z)^*  f^{\hat d_m}_{L d_j} (z) 
+ f^{\hat d_\ell}_{L d_j'} (z)^*  f^{\hat d_m}_{L d_j'} (z)  \cr
\noalign{\kern 5pt}
&\hskip 2.cm
+ f^{\hat d_\ell}_{L D_j^+} (z)^*  f^{\hat d_m}_{L D_j^+} (z) 
+ f^{\hat d_\ell}_{L D_j^-} (z)^*  f^{\hat d_m}_{L D_j^-} (z)  \Big\} + (L \go R) \Big] ~.
\label{photoncoupling4D2}
\end{align}
By making use of orthonormality relations, the $z$ integration can be done to lead to
\begin{align}
{\cal L}_\gamma^{d=4} &=  - i e A_\mu^\gamma (x) \sum_{j=1}^3 \bigg\{ \frac{2}{3} \,
\bar{\hat u}_{j} (x) \Gamma^\mu \hat u_{j} (x) 
- \frac{1}{3} \, \bar{\hat d}_{j} (x) \Gamma^\mu \hat d_{j}  (x)
\bigg\} ~, \cr
\noalign{\kern 5pt}
e &= g_w \sin \theta_W^0 ~, ~~
\sin \theta_W^0 = \frac{s_\phi}{\sqrt{\smash[b]{1+s_\phi^2}}} ~.
\label{photoncoupling4D3}
\end{align}

$Z$ couplings are given by
\begin{align}
{\cal L}_Z^{d=4} &= - ig_A \frac{\sqrt{\smash[b]{1+s_\phi^2}}}{\sqrt{2}} \, 
\int_1^{z_L} \frac{dz}{k} \cr
\noalign{\kern 5pt}
&\times \bigg\{
\mathring{Z}_\mu  \bigg[ \frac{1+c_H}{2}  \Big( \vec{\bar{\tilde u}} \,  \Gamma^\mu \, \vec{\tilde u}
- \vec{\bar{\tilde d}} \,  \Gamma^\mu \, \vec{\tilde d} \, \Big)
+ \frac{1- c_H}{2} \Big( \vec{\bar{\tilde u}}' \,  \Gamma^\mu \, \vec{\tilde u}' 
-  \vec{\bar{\tilde d}}' \,  \Gamma^\mu \, \vec{\tilde d}'  \, \Big)     \bigg] \cr
\noalign{\kern 5pt}
&\hskip 1.5cm
- \mathring{Z}_\mu^S \,  i \frac{s_H}{2} \Big[  \vec{\bar{\tilde u}} \,  \Gamma^\mu \, \vec{\tilde u}'
- \vec{\bar{\tilde u}}' \,  \Gamma^\mu \, \vec{\tilde u} 
- \vec{\bar{\tilde d}} \,  \Gamma^\mu \, \vec{\tilde d}'
+\vec{\bar{\tilde d}}' \,  \Gamma^\mu \, \vec{\tilde d} ~ \Big] \bigg\} \cr
\noalign{\kern 5pt}
&+ i g_A  \frac{\sqrt{2} s_\phi^2}{\sqrt{\smash[b]{1+s_\phi^2}}} 
\int_1^{z_L} \frac{dz}{k} \, \mathring{Z}_\mu \, J_\gamma^\mu   
\label{Zcoupling4D2}
\end{align}
where $J_\gamma^\mu $ is given in (\ref{photoncoupling4D2}).
Let us denote  $Z$ couplings of quarks as
\begin{align}
{\cal L}_Z^{d=4} &=- \frac{i}{\cos \theta_W^0 } Z_\mu \bigg\{ 
\sum_{j} \Big( 
g_{L u_j u_j}^Z \, \bar{\hat u}_{jL} \Gamma^\mu \hat u_{jL} + 
g_{R u_j u_j}^Z \, \bar{\hat u}_{jR} \Gamma^\mu \hat u_{jR} \Big)  \cr
\noalign{\kern 5pt}
&\hskip 1.7cm
+ \sum_{j,k} \Big( 
g_{L d_j d_k}^Z \, \bar{\hat d}_{jL} \Gamma^\mu \hat d_{kL} + 
g_{R d_j d_k}^W \, \bar{\hat d}_{jR} \Gamma^\mu \hat d_{kR} \Big) \bigg\} ~.
\label{Zcoupling4D3}
\end{align}
The couplings of up-type quarks are diagonal in flavor, but there appear off-diagonal couplings (FCNC)
for down-type quarks.
Insertion of (\ref{tower1}),   (\ref{notationUp1}) and (\ref{notationDown1}) into (\ref{Zcoupling4D2})
leads to 
\begin{align}
&g_{L u_j u_j}^Z  = g_w \frac{\sqrt{2kL}}{\sqrt{r_Z}} \int_1^{z_L} dz   \cr
\noalign{\kern 5pt}
&
\times \bigg\{ C(z, \lambda_Z) \Big(
\frac{1+c_H}{4}   f^{\hat u_j}_{L u_j} (z)^*  f^{\hat u_j}_{L u_j} (z) 
+ \frac{1-c_H}{4}   f^{\hat u_j}_{L u_j'} (z)^*  f^{\hat u_j}_{L u_j'} (z)   \cr
\noalign{\kern 5pt}
&\hskip 2.5cm
- \frac{2}{3}  \sin^2 \theta_W^0 \big[  f^{\hat u_j}_{L u_j} (z)^*  f^{\hat u_j}_{L u_j} (z) 
+  f^{\hat u_j}_{L u_j'} (z)^*  f^{\hat u_j}_{L u_j'} (z)  \big] \Big) \cr
\noalign{\kern 5pt}
&\hskip 1.0cm 
- i  \frac{s_H}{2} \hat S(z, \lambda_Z)   \big[  f^{\hat u_j}_{L u_j} (z)^*   f^{\hat u_j}_{L u_j'} (z) 
- f^{\hat u_j}_{L u_j'} (z)^*   f^{\hat u_j}_{L u_j} (z) \big] \bigg\} ~, \cr
\noalign{\kern 5pt}
&g_{L d_j d_k}^Z  = g_w \frac{\sqrt{2kL}}{\sqrt{r_Z}} \int_1^{z_L} dz  \sum_{\ell=1}^3 \cr
\noalign{\kern 5pt}
&
\times \bigg\{ C(z, \lambda_Z) \Big(
- \frac{1+c_H}{4}   f^{\hat d_j}_{L d_\ell} (z)^*  f^{\hat d_k}_{L d_\ell} (z) 
- \frac{1-c_H}{4}   f^{\hat d_j}_{L d_\ell'} (z)^*  f^{\hat d_k}_{L d_\ell'} (z)   \cr
\noalign{\kern 5pt}
&\hskip 2.5cm 
+ \frac{1}{3}  \sin^2 \theta_W^0 \big[  f^{\hat d_j}_{L d_\ell} (z)^*  f^{\hat d_k}_{L d_\ell} (z) 
+  f^{\hat d_j}_{L d_\ell'} (z)^*  f^{\hat d_k}_{L d_\ell'} (z)  \cr
\noalign{\kern 5pt}
&\hskip 4.0cm
+ f^{\hat d_j}_{L D_\ell^+} (z)^*  f^{\hat d_k}_{L D_\ell^+} (z) 
+ f^{\hat d_j}_{L D_\ell^-} (z)^*  f^{\hat d_k}_{L D_\ell^-} (z)  \big] \Big) \cr
\noalign{\kern 5pt}
&\hskip 1.0cm 
+ i  \frac{s_H}{2} \hat S(z, \lambda_Z)   \big[  f^{\hat d_j}_{L d_\ell} (z)^*   f^{\hat d_k}_{L d_\ell} (z) 
- f^{\hat d_j}_{L d_\ell'} (z)^*   f^{\hat d_k}_{L d_\ell} (z) \big] \bigg\} ~.
\label{Zcoupling4D4}
\end{align}
Formulas for $g_{R u_j u_j}^Z$ and $g_{R d_j d_k}^Z$ are obtained by the replacement $L \go R$
in each expression.

The $Z$ couplings  of down-type quarks are written in the matrix form;
$(\widehat g^Z_{L\,d})_{jk} = g_{L d_j d_k}^Z$ and 
$(\widehat g^Z_{R\, d})_{jk} = g_{R d_j d_k}^Z$.  One find for the two sets of parameters
in Table \ref{Tab:cDmu};
\begin{align}
\hbox{(a) ~} &\theta_H = 0.10 ~: \cr
\noalign{\kern 5pt}
&\begin{pmatrix} g^Z_{Luu} \cr g^Z_{Lcc} \cr g^Z_{Ltt} \end{pmatrix} = 
\begin{pmatrix} 0.3451 \cr 0.3451 \cr 0.3455 \end{pmatrix} g_w ~, ~~
\begin{pmatrix} g^Z_{Ruu} \cr g^Z_{Rcc} \cr g^Z_{Rtt} \end{pmatrix} = 
\begin{pmatrix} - 0.1538 \cr - 0.1538 \cr - 0.1534 \end{pmatrix} g_w ~,
\cr
\noalign{\kern 5pt}
&\widehat g^Z_{L\,d} = g_w  \begin{pmatrix} -0.4220 &  -3 \times 10^{-7}& -4 \times 10^{-9} \cr
-3 \times 10^{-7} &-0.4220  & -1 \times 10^{-7}  \cr 
 -4 \times 10^{-9}& -1 \times 10^{-7} &-0.4220 \end{pmatrix} , \cr
\noalign{\kern 5pt}
&\widehat g^Z_{R\,d} = g_w \begin{pmatrix} 0.0769 & -6  \times 10^{-7} & -4 \times 10^{-7} \cr
-6 \times 10^{-7} & 0.0769 & -3\times 10^{-6} \cr
-4 \times 10^{-7} &- 3 \times 10^{-6} & 0.0769 \end{pmatrix} , \cr
\noalign{\kern 10pt}
\hbox{(b) ~} &\theta_H = 0.15 ~: \cr
\noalign{\kern 5pt}
&\begin{pmatrix} g^Z_{Luu} \cr g^Z_{Lcc} \cr g^Z_{Ltt} \end{pmatrix} = 
\begin{pmatrix} 0.3441 \cr 0.3441 \cr 0.3449 \end{pmatrix} g_w ~, ~~
\begin{pmatrix} g^Z_{Ruu} \cr g^Z_{Rcc} \cr g^Z_{Rtt} \end{pmatrix} = 
\begin{pmatrix} - 0.1533 \cr - 0.1533 \cr - 0.1524 \end{pmatrix} g_w ~,
 \cr
\noalign{\kern 5pt}
&\widehat g^Z_{L\,d} = g_w  \begin{pmatrix} -0.4208 &  -7 \times 10^{-7}& -1 \times 10^{-8} \cr
-7 \times 10^{-7} &-0.4208  & -4 \times 10^{-7}  \cr 
 -1 \times 10^{-8}& -4 \times 10^{-7} &-0.4207 \end{pmatrix} , \cr
\noalign{\kern 5pt}
&\widehat g^Z_{R\,d} = g_w \begin{pmatrix} 0.0767 & -1  \times 10^{-6} & -1 \times 10^{-6} \cr
-1 \times 10^{-6} & 0.0767 & -7 \times 10^{-6} \cr
-1 \times 10^{-6} &- 7 \times 10^{-6} & 0.0767 \end{pmatrix} .
\label{Zcoupling4D5}
\end{align}
Although flavor-changing neutral currents (FCNC's) emerge for the down-type quarks, 
their magnitude is naturally suppressed.
FCNC's induce the mixing of neutral mesons ($M = K, B_d, B_s$)   at the tree level,  
yielding  $\Delta m_M \sim (m_M f_M^2/3 m_Z^2) (\widehat  g^Z_{d}|_{M})^2$ 
where $m_M$ and $f_M$ are the meson mass and
decay constant and $\widehat  g^Z_{d}|_{M}$ is the relevant coupling in $\widehat  g^Z_{L\,d}$ or
$\widehat  g^Z_{R\,d}$.
Making use of $(m_K, m_{B_d}, m_{B_s}) \sim  (0.498, 5.280, 5.367)\,$GeV and
$(f_K, f_{B_d}, f_{B_s}) \sim (0.156, 0.191, 0.274)\,$GeV, 
one finds, for $\theta_H = 0.10$, 
$(\Delta m_K, \Delta m_{B_d}, \Delta m_{B_s}) \sim (7 \times 10^{-20}, 5 \times 10^{-19}, 
6 \times 10^{-17})\,$GeV, which are much smaller than the experimental values
$(3.48 \times 10^{-15}, 3.36 \times 10^{-13}, 1.17 \times 10^{-11})\,$GeV.\cite{FCNCdata, FCNCanalysis}

The gauge invariance guarantees natural suppression of FCNC interactions.
This should be contrasted to the previous approaches of refs.\ \cite{Adachi2010, Cacciapaglia2008},
in which only $SU(3)_C \times SU(2)_L \times U(1)_Y$ invariance is imposed on the brane.
The requirement of the gauge invariance under ${\cal G} = SU(3)_C \times SO(5) \times U(1)_X$ 
restricts the form of brane interactions to (\ref{BraneInt1}), which yield the specific
mass terms of the form (\ref{braneFmass1}).  The resultant FCNC's are suppressed 
with a factor of order $(m_q/m_D)^2$ ($m_q= m_d, m_s, m_b$) as anticipated
from the effective theory developed in section 4.
The orbifold boundary condition breaks $SO(5)$ to $SO(4)$ so that one might expect
only ${\cal G}' = SU(3)_C \times SO(4) \times U(1)_X$ invariance on the brane.
As explained earlier, the above conclusion remains valid even with the ${\cal G}'$ gauge 
invariance alone being imposed.

We remark that the relative couplings  to $g_{L\,  \rm lepton}^W$  are
\begin{align}
\frac{1}{g_{L\,  \rm lepton}^W} 
\begin{pmatrix} g^Z_{Luu} \cr g^Z_{Ruu}\cr g^Z_{Ldd} \cr g^Z_{Rdd} \end{pmatrix} = 
\begin{pmatrix} 0.34588\cr  -0.15413  \cr -0.42295  \cr 0.07707 \end{pmatrix}, 
\begin{pmatrix} 0.34591 \cr -0.15411  \cr -0.42298 \cr 0.07706 \end{pmatrix} 
\label{relativeZcoupling}
\end{align}
for $\theta_H = 0.10$,  and $0.15$, respectively. 
The values in the SM with $\sin^2 \theta_W = 0.2312$ are 
$0.3458, -0.1541, -0.4229$ and $0.0771$.  The values (\ref{relativeZcoupling}) 
in the gauge-Higgs unification are very close to those in the SM.

\subsection{Yukawa couplings}

The flavor mixing in the down-type quarks  induces flavor-changing Yukawa couplings.
We show that its effect is extremely tiny.
The 4D Higgs field $H(x)$ is contained in $A_z^{\hat 4}$ in  the expansion (\ref{SO5pot1});
\begin{align}
&\tilde A_z = \mathring{H} (x,z) T^{\hat 4} + \cdots ,  \cr
\noalign{\kern 5pt}
&\mathring{H} (x,z)= \frac{1}{\sqrt{k}} \, H(x) h_H(z) + \cdots, \quad
h_H(z) = \sqrt{\frac{2}{z_L^2 -1}} \, z ~.
\label{Higgs1}
\end{align}
Inserting (\ref{Higgs1}) into the gauge interaction part of the action, one obtains 
\begin{align}
&-i g_A \int_1^{z_L} dz  \, \mathring{H} \sum_{\alpha = 1}^3   
\overline{\tilde{\check{\Psi}}}{}_{\bf (3,4)}^\alpha   \Gamma^5   \, T^{\hat 4} 
\, \tilde{\check{\Psi}}_{\bf (3,4)}^\alpha \cr
\noalign{\kern 5pt}
&= - \frac{g_A}{2\sqrt{2}} \int_1^{z_L} dz \, \mathring{H} \Big\{
\vec{\tilde u}{}^\dagger_L \, \vec{\tilde u}{}'_R + \vec{\tilde u}{}^{\prime\dagger}_L \, \vec{\tilde u}{}_R
+ \vec{\tilde u}{}^\dagger_R \, \vec{\tilde u}{}'_L + \vec{\tilde u}{}^{\prime\dagger}_R \, \vec{\tilde u}{}_L \cr
\noalign{\kern 5pt}
&\hskip 3.3cm
+\vec{\tilde d}{}^\dagger_L \, \vec{\tilde d}{}'_R + \vec{\tilde d}{}^{\prime\dagger}_L \, \vec{\tilde d}{}_R
+ \vec{\tilde d}{}^\dagger_R \, \vec{\tilde d}{}'_L + \vec{\tilde d}{}^{\prime\dagger}_R \, \vec{\tilde d}{}_L \Big\}
\label{Yukawa1}
\end{align}
where the notation (\ref{notation2}) has been used.
We insert (\ref{notationUp1}) and (\ref{notationDown1}) into (\ref{Yukawa1}) and integrate
over $z$.  In terms of mass eigenstates (\ref{notation3}) the Yukawa interactions are written as
\begin{align}
&- i H(x) \bigg\{ \sum_{j=1}^3  y_{u_j u_j}\Big( \hat u_{jL}^\dagger \hat u_{jR} -  \hat u_{jR}^\dagger \hat u_{jL} \Big) 
+ \sum_{j,k=1}^3  y_{d_j d_k} \Big( \hat d_{jL}^\dagger \hat d_{kR} - \hat d_{kR}^\dagger \hat d_{jL} \Big) \bigg\}
\label{Yukawa2}
\end{align}
where the Yukawa couplings are given by
\begin{align}
&y_{u_j u_j} = -i \frac{g_w \sqrt{kL}}{2\sqrt{2}}  \int_1^{z_L} dz \, h_H(z) \Big\{
f^{\hat u_j}_{L u_j} (z)^*  f^{\hat u_j}_{R u_j'} (z)  +  f^{\hat u_j}_{L u_j'} (z)^*  f^{\hat u_j}_{R u_j} (z)  \Big\}, \cr
\noalign{\kern 5pt}
&y_{d_j d_k} = -i \frac{g_w \sqrt{kL}}{2\sqrt{2}}  \int_1^{z_L} dz \, h_H(z) \sum_{m=1}^3 \Big\{
f^{\hat d_j}_{L d_m} (z)^*  f^{\hat d_k}_{R d_m'} (z)  +  f^{\hat d_j}_{L d_m'} (z)^*  f^{\hat d_k}_{R d_m} (z) \Big\} .
\label{Yukawa3}
\end{align}
Note that the Yukawa couplings in the up-type quark sector are diagonal in the generation space,
whereas those in the down-type quark sector have nonvanishing off-diagonal elements.

For the two sets of parameters in Table \ref{Tab:cDmu} one finds 
\begin{align}
\hbox{(a) ~} &\theta_H = 0.10 ~: \cr
\noalign{\kern 5pt}
&(y_{uu}, y_{cc}, y_{tt}) = ( 8.1376 \times 10^{-5},  2.5186 \times 10^{-3}, 0.69693) ~, \cr
\noalign{\kern 5pt}
&\widehat{y}_d = \begin{pmatrix} 1.1800 \times 10^{-5} & -1 \times 10^{-16} &  -2 \times 10^{-13} \cr
 -2 \times 10^{-18} & 2.2378 \times 10^{-4} &1 \times 10^{-11} \cr
 -9 \times 10^{-17} & 4 \times 10^{-13} & 1.1759 \times 10^{-2} \end{pmatrix}. \cr
\noalign{\kern 10pt}
\hbox{(b) ~} &\theta_H = 0.15 ~: \cr
\noalign{\kern 5pt}
&(y_{uu}, y_{cc}, y_{tt}) = ( 8.1222 \times 10^{-5},  2.5138 \times 10^{-3}, 0.69620) ~, \cr
&\widehat{y}_d = \begin{pmatrix} 1.1777 \times 10^{-5} & -3 \times 10^{-16} &  -9 \times 10^{-13} \cr
 -6 \times 10^{-18} & 2.2336 \times 10^{-4} &2 \times 10^{-11} \cr
 -4 \times 10^{-16} & 8 \times 10^{-13} & 1.1737 \times 10^{-2} \end{pmatrix}. 
\label{YukawaCoupling1}
\end{align}
Here $(\widehat{y}_d)_{jk} = y_{d_j d_k}$.  Note that in the evaluation we have used the values 
$(m_u, m_d) = (20,  2.90)\,$MeV for the reason described earlier.
The flavor-changing Yukawa couplings are exceedingly small.
Splitting of mass $\Delta m_M$ of neutral mesons ($M = \bar d_j d_k, d_j \bar d_k, j\not= k$)  due to $y_{d_j d_k}$ 
is estimated to be at most $[m_M/(m_{d_j} + m_{d_k}) ]^2  (m_M f_M^2/ m_H^2) ( y_{d_j d_k})^2$,\cite{FCNCanalysis}
which is much smaller than the observed $\Delta m_M$.

The values of the diagonal part of the Yukawa couplings can be understood from the effective theory as well.
Recalling that the 4D Higgs field $H(x)$ is the fluctuation mode of the AB phase $\theta_H$, the effective
interactions of $W, Z$ and fermion field $\psi_f$ with the Higgs field can be written as\cite{HK2009}
\begin{align}
&{\cal L} \sim - \overline{m}_W(\hat \theta_H)^2 W_\mu^\dagger W^\mu
 - \frac{1}{2} \overline{m}_Z(\hat \theta_H)^2 Z_\mu  Z^\mu 
 - \overline{m}_f (\hat \theta_H) \bar \psi_f \psi_f ~, \cr
\noalign{\kern 5pt}
&\hat \theta_H (x) = \theta_H + \frac{H(x)}{f_H} ~.
\label{effectiveHiggsInt1}
\end{align}
The mass functions are, in good approximation, given by
\begin{align}
&\overline{m}_W(\hat \theta_H) \sim  a_W \sin \hat \theta_H ~, \cr
\noalign{\kern 5pt}
&\overline{m}_Z(\hat \theta_H) \sim  a_Z \sin \hat \theta_H ~, \cr
\noalign{\kern 5pt}
&\overline{m}_f (\hat \theta_H)  \sim  \begin{cases} a_f \sin \hat \theta_H  &\hbox{in the A model} \cr
\noalign{\kern 5pt}
a_f \sin \onehalf \hat \theta_H  &\hbox{in the B model} ~, \end{cases}
\label{effectiveHiggsInt2}
\end{align}
where $a_W$, $a_Z$ and $a_f$ are constants.
At the tree level $m_W = \overline{m}_W(\theta_H) = \onehalf g_w f_H \sin \theta_H$,
$m_Z =  \overline{m}_Z(\theta_H) = m_W/\cos \theta_W^0$ and $m_f = \overline{m}_f(\theta_H)$.
Expanding the mass functions in (\ref{effectiveHiggsInt1}) around $\theta_H$, one find the Higgs
couplings to be
\begin{align}
&g_{WWH} =  \frac{2 m_W^2 \cos\theta_H}{f_H \sin\theta_H} = g_w m_W \cos \theta_H ~, \cr
\noalign{\kern 5pt}
&g_{ZZH} =  \frac{2 m_Z^2 \cos\theta_H}{f_H \sin\theta_H} = \frac{g_w m_Z}{\cos \theta_W^0} \, \cos \theta_H ~, \cr
\noalign{\kern 5pt}
&y_f = \begin{cases} \myfrac{m_f \cos\theta_H}{f_H \sin\theta_H} 
= \myfrac{m_f }{v_\SM} \, \cos\theta_H &\hbox{in the A model} \cr
\noalign{\kern 5pt}
\myfrac{m_f \cos \onehalf \theta_H}{2 f_H \sin \onehalf \theta_H} 
= \myfrac{m_f }{v_\SM}  \, \cos^2  \onehalf \theta_H &\hbox{in the B model}  ~. \end{cases} 
\label{effectiveHiggsInt3}
\end{align}
Here $v_\SM= f_H \sin\theta_H$.  In other words, compared to the couplings in the SM,
the Higgs couplings of $W$ and $Z$ in the gauge-Higgs unification are suppressed by a factor $\cos \theta_H$.
The Yukawa couplings of quarks and leptons are suppressed by a factor  $\cos \theta_H$ in the A model
and by a factor  $\cos^2 \onehalf \theta_H$ in the B model.

The diagonal part of the evaluated Yukawa couplings (\ref{YukawaCoupling1}) are well described
by the formula in (\ref{effectiveHiggsInt3}).   
Denoting the couplings in the SM by $y_f^\SM = m_f/v_\SM$,  one finds
\begin{align}
\hbox{(a) ~} &\theta_H = 0.10 ~: ~ \cos^2 \onehalf \theta_H = 0.99750 \cr
\noalign{\kern 5pt}
&\bigg( \frac{y_{uu}}{y_u^\SM} ,  \frac{y_{cc}}{y_c^\SM} , \frac{y_{tt}}{y_t^\SM} \bigg) = 
(0.99758, 0.99758, 0.99826) ~, \cr
\noalign{\kern 5pt}
& \bigg( \frac{y_{dd}}{y_d^\SM} ,  \frac{y_{ss}}{y_s^\SM} , \frac{y_{bb}}{y_b^\SM}  \bigg) = 
( 0.99758,  0.99758,  0.99758) ~. \cr
\noalign{\kern 10pt}
\hbox{(b) ~} &\theta_H = 0.15 ~: ~ \cos^2 \onehalf \theta_H =  0.99439 \cr
\noalign{\kern 5pt}
& \bigg( \frac{y_{uu}}{y_u^\SM} ,  \frac{y_{cc}}{y_c^\SM} , \frac{y_{tt}}{y_t^\SM}  \bigg) = 
( 0.99456,  0.99456, 0.99607) ~, \cr
\noalign{\kern 5pt}
& \bigg( \frac{y_{dd}}{y_d^\SM} ,  \frac{y_{ss}}{y_s^\SM} , \frac{y_{bb}}{y_b^\SM}  \bigg) = 
(0.99456,  0.99456,  0.99456) ~. 
\label{YukawaCoupling2}
\end{align}
The deviation from the SM is rather small.

We would like to add a comment.
As explained in Section 2, the neutral physical scalar of  $\Phi_{({\bf 1}, {\bf 4})}$
has a large mass $(\gg m_\KK)$ so that its couplings to quarks and leptons at low energies
are negligible, playing no role in flavor changing  processes.

\section{Summary and discussions}

In this paper we have shown that the flavor mixing in the quark sector can be incorporated 
in the GUT inspired $SU(3)_C \times SO(5) \times U(1)_X$ gauge-Higgs unification.
The brane interactions on the UV brane are responsible both for splitting the mass spectrum
between the up-type quarks and down-type quarks and for generating flavor mixing
in the charged current  ($W$) interactions.  
Quite reasonable form of the CKM matrix has been obtained.
The mixing, in general, induces  FCNC interactions in the $Z$ couplings of quarks.  
It is shown that the FCNC 
interactions are naturally suppressed, with a suppression factor of order $10^{-6}$.
The suppression is a result of the $SU(3)_C \times SO(5) \times U(1)_X$ or
$SU(3)_C \times SO(4) \times U(1)_X$ 
gauge invariance which allows only a certain
class of interactions on the UV brane.  In addition to presenting rigorous evaluation
of the gauge couplings, we have also given an explanation in terms of the 
effective theory which illustrates how the natural suppression of the FCNC interactions
results in the gauge-Higgs unification.
The flavor-mixing induces flavor-changing Yukawa couplings as well.
We have confirmed that those couplings are extremely small.

There remains an issue to be clarified.
In the present model we could obtain a consistent spectrum and mixing only
if the up quark mass $m_u$ were larger than the down quark mass $m_d$.  
With the minimal matter content in the GUT inspired  gauge-Higgs unification,
$m_d$ necessarily becomes smaller than $m_u$.  
One may have an additional field which affects $m_u$, or may consider
the running of quark masses which reverses the order of $m_u$ and $m_d$
at low energies.  We leave the issue for future investigation.

In the GUT inspired  gauge-Higgs unification we have chosen negative bulk mass
parameters.  With positive bulk mass parameters there arise exotic light fermions
with the same quantum numbers as the down-type quarks.
Although negative  bulk mass parameters imply that left-handed (right-handed)
light quarks are localized near the IR (UV) brane, we have shown that 
the $W$ and $Z$ couplings of all quarks are very close to those in the SM.
This is one of the remarkable properties in the gauge-Higgs unification 
in the RS space.
Similarly negative bulk mass parameters of leptons are preferred to positive ones,
as positive ones yield additional light neutral  fermions.

The sign of the bulk mass parameters of quarks and leptons can be investigated
by $e^+ e^-$ collider experiments, as the couplings of quarks and leptons to
$Z'$ bosons, namely KK excited states of $Z$, $\gamma$ and $Z_R$,  have
large parity violation.
It has been shown in the previous A-model of $SO(5) \times U(1)$ gauge-Higgs
unification that right-handed quarks and leptons have much larger couplings
to $Z'$ bosons so that in the process $e^+ e^- \go \mu^+ \mu^-$, for instance, 
significant deviation from the SM appears even at 250$\,$GeV ILC with 250$\,{\rm fb}^{-1}$
data.  If the $e^-$ beam is polarized in the left-handed mode, there would be no
deviation from the SM, whereas, if the $e^-$ beam is polarized in the right-handed mode,
then there appears large deviation.  By changing the polarization of the $e^-$ beam,
one can see a distinct pattern of deviation.  Similar effects are seen in the 
forward-backward asymmetry in various processes as well.
In the present B-model left-handed leptons and quarks have much larger couplings
to $Z'$ bosons than right-handed ones.  As a consequence the pattern of the dependence
on the $e^-$ polarization is reversed in comparison with that in the A-model.
ILC experiments can provide rich information on underlying physics.

Gauge-Higgs unification is formulated in five or higher dimensions in which the running of
gauge couplings is much more rapid than in four dimensions.\cite{Yamatsu2016} 
In this paper we have analyzed the $W$ and $Z$ couplings of quarks below the 
KK mass scale $m_\KK$.  All relations presented in this paper should be understood 
as those for the energy scale below $m_\KK$.  Above $m_\KK$ effects of KK modes
need to be properly incorporated.
Gauge-Higgs unification is a new approach to physics beyond the SM.  
It may provide a key to solve the problems of dark matter, gauge hierarchy, 
neutrinos, Higgs couplings, and grand unification as well.\cite{Burdman2003}-\cite{Yamada2019}
We will come back to these issues in the future.

\section*{Acknowledgements}

This work was supported in part 
by European Regional Development Fund-Project Engineering Applications of 
Microworld Physics (No.\ CZ.02.1.01/0.0/0.0/16-019/0000766) (Y.O.), 
by the National Natural Science Foundation of China (Grant Nos.~11775092, 
11675061, 11521064, 11435003 and 11947213) (S.F.), 
by the International Postdoctoral Exchange Fellowship Program (IPEFP) (S.F.), 
and by Japan Society for the Promotion of Science, 
 Grants-in-Aid  for Scientific Research,  No.\ 19K03873 (Y.H.) and Nos.\  18H05543 and 19K23440 (N.Y.).

\appendix

\section{Basis functions}

We summarize basis functions in the RS space.  
We define
\begin{align}
&F_{\alpha,\beta}(u,v) \equiv J_\alpha(u) Y_\beta(v) - Y_\alpha(u)J_\beta(v)
\end{align}
where $J_\alpha(x)$ and $Y_\alpha(x)$
are Bessel functions of the 1st and 2nd kind, respectively.
For gauge bosons $C = C(z;\lambda)$ and $S = S(z;\lambda)$ are defined by
\begin{align}
C(z;\lambda) &= + \frac{\pi}{2} \lambda z z_L F_{1,0}(\lambda z, \lambda z_L), \cr
C'(z;\lambda) &= + \frac{\pi}{2} \lambda^2 z z_L F_{0,0}(\lambda z, \lambda z_L), \cr
S(z;\lambda) &= - \frac{\pi}{2} \lambda z F_{1,1}(\lambda z, \lambda z_L), \cr
S'(z;\lambda) &= - \frac{\pi}{2} \lambda^2 z F_{0,1}(\lambda z, \lambda z_L). 
\label{gaugeF1}
\end{align}
We note that $C S' - SC' = \lambda z$.

For massless fermions we define
\begin{align}
\begin{pmatrix}C_R \cr S_R \end{pmatrix} (z;\lambda,c) 
&= \mp \frac{\pi}{2} \lambda \sqrt{z z_L} \, F_{c-\frac{1}{2},c \pm \frac{1}{2}}(\lambda z,\,\lambda z_L) ~, \cr 
\noalign{\kern 5pt}
\begin{pmatrix} C_L \cr S_L \end{pmatrix} (z;\lambda,c)   
&= \pm\frac{\pi}{2} \lambda \sqrt{z z_L} \, F_{c+\frac{1}{2},c \mp \frac{1}{2}}(\lambda z,\,\lambda z_L) ~.
\label{fermionF1}
\end{align}
These satisfy $ C_L C_R - S_L S_R = 1$,  $C_L(z;\lambda, -c) = C_R(z; \lambda, c)$,
and $S_L(z;\lambda, -c) = - S_R(z; \lambda, c)$.
For massive fermions such as $D^\pm$ fields with $m_D \not= 0$ we define basis functions
\begin{align}
\begin{pmatrix} {\cal C}_{R1} \cr {\cal C}_{L1} \end{pmatrix}(z; \lambda, c, \tilde m) 
&= \begin{pmatrix}C_R \cr C_L \end{pmatrix} (z; \lambda, c+\tilde{m})
   +\begin{pmatrix}C_R \cr C_L \end{pmatrix} (z; \lambda, c-\tilde{m}) ~, \cr
\noalign{\kern 5pt}
\begin{pmatrix}{\cal C}_{R2} \cr {\cal C}_{L2} \end{pmatrix} (z; \lambda, c, \tilde m) 
&= \begin{pmatrix} S_R \cr S_L \end{pmatrix} (z; \lambda, c+\tilde{m})
   - \begin{pmatrix} S_R \cr S_L \end{pmatrix} (z; \lambda,c-\tilde{m}) ~, \cr
\noalign{\kern 5pt}
\begin{pmatrix} {\cal S}_{R1} \cr {\cal S}_{L1} \end{pmatrix} (z; \lambda, c, \tilde m) 
&= \begin{pmatrix} S_R \cr S_L \end{pmatrix} (z; \lambda, c+\tilde{m})
   +\begin{pmatrix} S_R \cr S_L \end{pmatrix} (z; \lambda,c-\tilde{m}) ~, \cr
\noalign{\kern 5pt}
\begin{pmatrix} {\cal S}_{R2}  \cr {\cal S}_{L2} \end{pmatrix} (z; \lambda, c, \tilde m) 
&= \begin{pmatrix} C_R \cr C_L \end{pmatrix} (z; \lambda, c+\tilde{m})
   - \begin{pmatrix} C_R \cr C_L \end{pmatrix} (z; \lambda, c-\tilde{m}) ~.
\label{MfermionBasis1}
\end{align}
These functions satisfy various relations which are summarized in Appendix B of ref.\  \cite{FHHOY2019a}.

\vskip 1.cm

\def\jnl#1#2#3#4{{#1}{\bf #2},  #3 (#4)}

\def\Zphys{{\em Z.\ Phys.} }
\def\jssc{{\em J.\ Solid State Chem.\ }}
\def\jpsJ{{\em J.\ Phys.\ Soc.\ Japan }}
\def\ptps{{\em Prog.\ Theoret.\ Phys.\ Suppl.\ }}
\def\PTP{{\em Prog.\ Theoret.\ Phys.\  }}
\def\PTEP{{\em Prog.\ Theoret.\ Exp.\  Phys.\  }}
\def\JMP{{\em J. Math.\ Phys.} }
\def\NPB{{\em Nucl.\ Phys.} B}
\def\NP{{\em Nucl.\ Phys.} }
\def\PLB{{\it Phys.\ Lett.} B}
\def\PL{{\em Phys.\ Lett.} }
\def\PRL{\em Phys.\ Rev.\ Lett. }
\def\PRB{{\em Phys.\ Rev.} B}
\def\PRD{{\em Phys.\ Rev.} D}
\def\PRe{{\em Phys.\ Rep.} }
\def\AP{{\em Ann.\ Phys.\ (N.Y.)} }
\def\RMP{{\em Rev.\ Mod.\ Phys.} }
\def\ZPC{{\em Z.\ Phys.} C}
\def\SCI{\em Science}
\def\CMP{\em Comm.\ Math.\ Phys. }
\def\MPLA{{\em Mod.\ Phys.\ Lett.} A}
\def\IJMPA{{\em Int.\ J.\ Mod.\ Phys.} A}
\def\IJMPB{{\em Int.\ J.\ Mod.\ Phys.} B}
\def\EPJC{{\em Eur.\ Phys.\ J.} C}
\def\PR{{\em Phys.\ Rev.} }
\def\JHEP{{\em JHEP} }
\def\JCAP{{\em JCAP} }
\def\cmp{{\em Com.\ Math.\ Phys.}}
\def\JPA{{\em J.\  Phys.} A}
\def\JPG{{\em J.\  Phys.} G}
\def\NJP{{\em New.\ J.\  Phys.} }
\def\CQG{\em Class.\ Quant.\ Grav. }
\def\ATMP{{\em Adv.\ Theoret.\ Math.\ Phys.} }
\def\ibid{{\em ibid.} }
\def\ChP{{\em Chin.Phys.}C}


\renewenvironment{thebibliography}[1]
         {\begin{list}{[$\,$\arabic{enumi}$\,$]}  
         {\usecounter{enumi}\setlength{\parsep}{0pt}
          \setlength{\itemsep}{0pt}  \renewcommand{\baselinestretch}{1.2}
          \settowidth
         {\labelwidth}{#1 ~ ~}\sloppy}}{\end{list}}

\leftline{\Large \bf References}


\end{document}